\documentstyle[12pt]{article}
\input epsf.tex
\begin{document}

\def\a{\alpha}
\def\b{\beta}
\def\ch{\chi}
\def\d{\delta}
\def\e{\epsilon}
\def\f{\phi}
\def\g{\gamma}
\def\h{\eta}
\def\i{\iota}
\def\j{\psi}
\def\k{\kappa}
\def\l{\lambda}
\def\m{\mu}
\def\n{\nu}
\def\o{\omega}
\def\p{\pi}
\def\q{\theta}
\def\r{\rho}
\def\s{\sigma}
\def\t{\tau}
\def\u{\upsilon}
\def\x{\xi}
\def\z{\zeta}
\def\D{\Delta}
\def\F{\Phi}
\def\G{\Gamma}
\def\J{\Psi}
\def\L{\Lambda}
\def\O{\Omega}
\def\P{\Pi}
\def\S{\Sigma}
\def\U{\Upsilon}
\def\X{\Xi}
\def\T{\Theta}

\def\Ab{\bar{A}}
\def\gi{g^{-1}}
\def\li{{ 1 \over \l } }
\def\lb{\l^{*}}
\def\zb{\bar{z}}
\def\ub{u^{*}}
\def\vb{v^{*}}
\def\Tb{\bar{T}}
\def\pp {\partial }
\def\pb {\bar{\partial }}
\def\be{\begin{equation}}
\def\ee{\end{equation}}
\def\ben{\begin{eqnarray}}
\def\een{\end{eqnarray}}

\hsize=16.5truecm
\addtolength{\topmargin}{-0.8in}
\addtolength{\textheight}{1in}
\hoffset=-.6in

\thispagestyle{empty}
\begin{flushright} \ April \ 1995\\
hep-th/9505017\\
SNUCTP 95-55\\
\end{flushright}
\begin{center}
 {\large\bf Classical Matrix sine-Gordon Theory  }\\[.1in]
\vglue .5in
 Q-Han Park\footnote{ E-mail address; qpark@nms.kyunghee.ac.kr }
\\[.2in]
{and}
\\[.2in]
H. J. Shin\footnote{ E-mail address; hjshin@nms.kyunghee.ac.kr }
\\[.2in]
{\it
Department of Physics \\
and \\
Research Institute of Basic Sciences \\
Kyunghee University\\
Seoul, 130-701, Korea}
\\[.2in]
{\bf ABSTRACT}\\[.2in]
\end{center}
\vglue .1in
The matrix sine-Gordon theory, a matrix generalization of the well-known
sine-Gordon theory, is studied. In particular, the $A_{3}$-generalization
where fields take value in $SU(2)$ describes integrable deformations of
conformal field theory corresponding to the coset $SU(2) \times SU(2) /SU(2)$.
Various classical aspects of the matrix sine-Gordon theory are addressed.
We find exact solutions, solitons and breathers which  generalize those of
the sine-Gordon theory with internal degrees of freedom, by applying the
Zakharov-Shabat dressing method and
explain their physical properties. Infinite current conservation laws and the
B\"{a}cklund transformation of the theory are obtained from the zero curvature
formalism of the equation of motion. From the B\"{a}cklund transformation,
we also derive exact solutions as well as a nonlinear superposition principle
by
making use of the Bianchi's permutability theorem.

\newpage
\section{Introduction}
The sine-Gordon theory is the most well-known example of relativistic
integrable field theories in 1+1 dimensions. Exact solutions, topological
solitons and breathers, are known and various applications of these solutions
have been made in the study of a wide range of physical systems. However,
in many cases, such an application has been made after a truncation of the
physical systems, i.e. suppressing all degrees of freedom except one scalar
field, so that the sine-Gordon theory becomes an effective description of
the reduced system. In this regard, it is desirable to have generalizations
of the sine-Gordon theory with additional degrees of freedom whereas the
integrability of the theory is maintained to provide exact solutions.
One such example is the so-called complex sine-Gordon
theory which appeared in the description of the relativistic vortex
motion in a superfluid\cite{lund}, in a treatment of O(4) nonlinear
sigma model\cite{pohl} and more recently in the context of conformal field
theory\cite{bakas}. In the complex sine-Gordon theory, the scalar field
is complex valued so that the phase factor becomes an additional physical
degree.
Other types of nonabelian generalization of the
sine-Gordon theory, carrying extra internal degrees, have been considered
recently
in the context of integrable
deformation of the coset conformal field theory\cite{park}, particularly in
the case of the critical Ising model deformed by $\F_{(2,1)}$ and
$\F_{(3,1)}$ operators\cite{shin}. In fact, to each classical Lie
algebras, there exist nonabelian generalizations of the
sine-Gordon theory  which admit a positive definite kinetic energy when
certain criteria of the sl(2) embedding are met\cite{hol}. The complex
sine-Gordon theory and the critical Ising model deformed by $\F_{(2,1)}$
operator
arise as first two examples in these generalizations which correspond to the
Lie
algebra $B_2$ and $A_3$ respectively.

In this Paper, we study in detail the $A_3$-generalization of the sine-Gordon
theory, which we call ``the matrix sine-Gordon theory", at the classical
level.\footnote{The name ``matrix sine-Gordon" has been also used for a
different
model on the group $SO(n)$ in \cite{Budagov}.}
At the quantum level, this theory may be regarded as an integrable
deformation of a minimal model\cite{goddard} corresponding to the coset
$SU_{N}(2) \times
SU_{N}(2) /SU_{2N}(2)$, in particular the deformation of the critical Ising
model
by the energy operator $\F_{(2,1)}$ if the level $N=1$.
We demonstrate the classical integrability of the theory by deriving
infinite current conservation laws from the zero curvature formalism of
the equation of motion. The B\"{a}cklund transformation is also obtained from
the zero curvature formalism,  and using the Bianchi's theorem of
permutability\cite{eisen} a nonlinear superposition principle is derived for
the solutions
of the matrix sine-Gordon theory. Exact solutions, solitons and breathers which
generalize
those of the sine-Gordon theory, are obtained by making
use of the Zakharov-Shabat dressing method\cite{zak}. These solutions carry
internal degrees
of freedom which affect the scattering process. Two soliton solutions for the
soliton(antisoliton) - soliton(antisoliton) scattering is given explicitly in
terms of
parameters $u$ and $\q $ which describe the relative velocity of solitons and
their relative
internal directions respectively.
It is shown that the nonabelian effect which is controlled by the parameter $\q
$ makes solitons
less repulsive for larger $\q $  while the scattering time of far removed
solitons does not
depend on $\q $.
An explicit form of nonabelian breather solution is also given in terms of both
energy
configuration and internal components with parameters $v$ and $\q $.
The breather solution is a bound state of soliton and antisoliton which
oscillates in time
with angular frequency $2\sqrt{-\k}v/\sqrt{1+v^2 }$.
As $\q $ increases, the breathing mode of the potential energy configuration
diminishes and
at $\q = \pi $, it becomes completely breathless! However, the internal
components still
oscillate at $\q = \pi$ which keeps the static potential energy configuration.
Alternative derivation of one and two soliton solutions are also given using
the B\"{a}cklund
transformation and the nonlinear superposition principle.

The plan of the Paper is the following; in Sec.2, we define the model in
terms of the gauged Wess-Zumino-Witten functional and address issues on
the gauge invariance and the gauge fixing.  We present the zero
curvature formalism of the equation of motion for the model and from which,
derive infinite current conservation laws thereby demonstrating the
integrability
of the model. In Sec.3,  exact solutions, N solitons and breathers, are derived
rigorously by making use of the dressing method. A pictorial description of
these
solutions is given and their physical meaning  is discussed.  Sec.4 deals with
the
B\"{a}cklund transformation and the nonlinear superposition principle
and Sec.5 is a discussion.

\section{The Model}
We introduce the matrix sine-Gordon theory from the context of conformal
field theory and its integrable deformations.  The action principle for the
G/H-coset conformal field theory may be given in terms of the gauged
Wess-Zumino-Witten(WZW) functional\cite{GWZW}, which in light-cone variables is
\be
S(g,A, \Ab ) = S_{WZW}(g) +
{1 \over 2\pi }\int \mbox {Tr} (- A\pb g \gi + \Ab \gi \pp g
 + Ag\Ab \gi - A\Ab )
\ee
and $S_{WZW}(g)$ is the action of group G WZW model
\be
S_{WZW}(g)=-{1\over 4\pi }\int_{\S }\mbox{Tr } \gi \pp g \gi \pb g
- {1 \over 12\pi }\int_{B}\mbox{Tr } \tilde{g}^{-1}d \tilde{g}\wedge
\tilde{g}^{-1}d \tilde{g}
\wedge \tilde{g}^{-1}d \tilde{g} \ .
\ee
$\tilde{g}$ is an extension of a map $g:\S \rightarrow G$ to a
three-dimensional
manifold $B$ with boundary $\S , \ \ \tilde{g}|_{\pp B }=g$, and
the connection fields $A, \Ab$ gauge the anomaly free subgroup H of G.
Here, we take the diagonal embedding of H in $G_{L} \times G_{R}$, where
$G_{L}$ and $G_{R}$ denote left and right group actions by multiplication
$(g \rightarrow g_{L}gg_{R}^{-1})$, so that Eq.(1) is invariant under the
vector gauge transformation; $ g \rightarrow hgh^{-1}$ with $h :  \S
\rightarrow $H. In particular, the minimal unitary series in conformal field
theory arise from the restriction to  the coset, $(SU(2)_{L} \times
SU(2)_{N}) / SU(2)_{L + N}$, where integers $L, N$ denote the level of the
Kac-Moody algebra\cite{goddard}.
In this case, the full theory is given formally by functional integrals,
\be
\int [dg_{1}][dg_{2}][dA][d\Ab ] \exp{ iI_{0}(g_{1}, g_{2}, A , \Ab ) }
\ee
where
\be
 I_{0}(g_{1}, g_{2}, A , \Ab ) =  LS_{WZW}(g_{1}, A, \Ab ) + NS_{WZW}(g_{2}, A,
\Ab  )
\ee
and $A, \Ab  $ gauge simultaneously the diagonal subgroups of $SU(2) \times
SU(2)$.

The matrix sine-Gordon theory is defined as a massive deformation of the
minimal
series with $L = N$ which, at the action level, is adding a potential term to
the action,
\be
I(g_{1},g_{2},A,\Ab ,\k ) = I_{0}(g_{1},g_{2},A,\Ab ) -
{N\k \over 2\pi }
\int \mbox{Tr} (g^{-1}_{1}g_{2} + g^{-1}_{2}g_{1}) ,
\ee
where $\k $ is a coupling constant.\footnote{Here, we asuume $g_{1}$ and
$g_{2}$
to take values in SU(2). The U(2) case has been considered in \cite{shin} .}
Note that the potential term is invariant
under the similarity transform; $g_1 \rightarrow s g_1 s^{-1} ,\ g_2
\rightarrow
 s g_2 s^{-1}$ so that the vector gauge invariance of the action is maintained.
In the convention of coset conformal field theory, the potential term
transforms at the classical level as (doublet, singlet) so that it corresponds
to the integrable perturbation of minimal series by the operator
$\Phi_{(2,1)}$\cite{zamol}. In this Paper, we focus only on the classical
aspect of the theory so that the level $N$ is irrelevant. In order to
understand
the vacuum structure, we parameterize $\gi_{1}g_{2}$ by
\be
\gi_{1}g_{2} =  \exp(i\f \hat{\s}) \ , \
\hat{\s} = \sum_{i=1}^{3} a_{i}\s_{i}
\ee
where $\s_{i}$ are Pauli matrices and the coefficients $a_i $ are normalized to
one,
$\sum_{i=1}^{3}a_{i}a_{i} = 1$.
Then, the potential becomes
\be
V={N\k \over 2\pi } \mbox{Tr} (g^{-1}_{1}g_{2} + g^{-1}_{2}g_{1})
= {2N\k \over \pi }\cos \f \ .
\ee
If the coupling constant $\k < 0$, $V$ possesses degenerate vacua at
$\f = 2n\pi $ for integer $ n $ so that $ \gi_{1}g_{2}=1$ or for any
arbitrary  $g_{1}=g_{2}$ valued in $SU(2)$. Note that a specific vacuum is
characterized
only by the integer $n$ independently of $a_i$. The degeneracy of the
vacuum allows soliton solutions which interpolate different vacuua.
The explicit solutions will be found in Sec.3 and Sec.4. The topological
soliton numbers are defined by the difference $\D n = n_1 - n_2 $ of
integer values of two interpolating vacuua.
If $\k > 0$,
degenerate vacuua occur at $\f = (2n+1)\pi $ for integer $ n $ so that
$ \gi_{1}g_{2}=-1$, or for any arbitrary  $g_{1}=-g_{2}$.
{}From now on, we will restrict ourselves to the $\k < 0$ case only. $\k > 0$
case will be discussed in Sec.5.

The classical equation of motion arising from the action Eq.(5) is
\ben
&& [\  \pp + \gi_{1} \pp g_{1} + \gi_{1} A g_{1} \ , \ \pb + \Ab \ ] - \k (
g^{-1}_{1}g_{2} - g^{-1}_{2}g_{1})  =   0  \nonumber \\
&& [\  \pp + \gi_{2} \pp g_{2} + \gi_{2} A g_{2} \ , \ \pb + \Ab \ ]  + \k
(g^{-1}_{1}g_{2} - g^{-1}_{2}g_{1})  = 0
\een
whereas variations of the action with respect to  $A$ and $\Ab $ give rise to
the constraint
equation,
\ben
 - \pb g_{1} \gi_{1} &+& g_{1}\Ab \gi_{1} - \pb g_{2} \gi_{2} + g_{2}\Ab
\gi_{2} - 2\Ab  = 0
\nonumber \\
 \gi_{1} \pp g_{1}  &+& \gi_{1} A g_{1} +  \gi_{2} \pp g_{2}  +
\gi_{2} A g_{2}- 2A = 0 \ .
\een
The constraint equation in Eq.(9), when combined with the equation
of motion in Eq.(8), results in the flatness
condition of $A$ and $ \Ab $,
\be
\pp \bar{A} - \pb A + [A, \bar{A}] =0,
\ee
which reflects the vector gauge invariance of the action. In the following,
we consider two types of different gauge fixing.
Assume that the underlying manifold $\S $ is the flat two-dimensional
Minkowski space $R^{1+1}$. Then the flatness of $A$
and $\Ab $ allows us to choose a {\it ``nonlocal gauge"}; \footnote{ For
general $\S$, such a nonlocal gauge is not always possible because
holonomies of $A, \Ab$ could be nontrivial.}
\be
  A=\Ab =0 .
\ee
The equation of motion in the nonlocal gauge becomes
\ben
&& \pb (\gi_{1} \pp g_{1} ) + \k ( g^{-1}_{1}g_{2} - g^{-1}_{2}g_{1})  =
0  \nonumber \\
&& \pb (\gi_{2} \pp g_{2} )-\k  (g^{-1}_{1}g_{2} - g^{-1}_{2}g_{1})  = 0
\een
whereas the constraint equation becomes
\be
  \pb g_{1} \gi_{1} + \pb g_{2} \gi_{2}   = 0\ , \ \gi_{1} \pp g_{1}  +
 \gi_{2} \pp g_{2} = 0 \ .
\ee
In the abelian limit, where $g_{1}=\exp{i\f_{1}}\s_{1} , \
g_{2}=\exp{i\f_{2}\s_{1}}$,
the constraint equation may be solved locally by $\f_{1} = \f = -\f_{2}$ so
that the equation of motion in terms of $\f$ becomes precisely the
sine-Gordon equation. However, for $g_{1}, g_{2}$ valued in $SU(2)$,
the constraint equation in general can not be solved locally. Consequently,
in the nonlocal gauge, a local parametrization solving the constraint is not
possible as in the abelian case.\footnote{Even though local parametrization
is not possible, exact solutions can be constructed explicitly as in Sec.3 and
Sec.4.}

Nevertheless, there exists another type of gauge fixing, so-called ``the
unitary gauge"\cite{wit}, which allows a local parametrization solving the
constraint
in the following sense; for any given $g_{1}$ and $g_{2}$, one may bring
$g_{2}$ into a form $g_{2}= \exp (i \f \s_3)$
via the similarity transform; $g_1 \rightarrow s g_1 s^{-1}, g_{2} \rightarrow
s g_2 s^{-1}$ for some $s$. Thus, the scalar function $\f$ parameterizes the
equivalence classes of $g_{2}$ with the equivalence relation given by the
similarity transform. The remaining $U(1)$ gauge symmetry which leaves
$g_{2}$ invariant may be used to fix $g_1$ to give {\it the unitary gauge};
\be
g_{1} = \pmatrix{u e^{ -i \f}& ~& i \sqrt{1 - u \ub } e^{ i \f} \cr
	i \sqrt{1 - u \ub } e^{-i \f } & ~& \ub  e^{ i \f} } \ , \
g_{2}=\pmatrix{ e^{ i \f } &  0 \cr
0 &  e^{ -i \f} }  \ .
\ee
In this gauge, the constraint equation can be solved explicitly for
$A$ and $\bar{A}$ in terms of $u$ and $\f$,
\be
A=\pmatrix{ a &  -b^* \cr b &  -a } \ , \ \ \
\bar{A} = \pmatrix{ \bar{a} & -\bar{b}^* \cr
\bar{b} & -\bar{a} }
\ee
where
\ben
b &=& {-i e^{-2i \f } \pp u \over 2 \sqrt{1 - u \ub }}
- { i\pp u + i\pp \ub  \over 8 \sqrt{1 - u \ub } \sin^2 \f } \nonumber \\
a &=& { (e^{2 i \f } \ub  - u) \pp \ub  -( e^{ -2i \f } u - \ub  )
\pp u \over 4 ( 1- u \ub ) } - { (u - \ub  ) (\pp u + \pp \ub  )
\over 16 ( 1- u \ub  ) \sin^2 \f },
\een
and
\ben
\bar{b} &=& {\pb \ub  \over 4 \sqrt{1 - u \ub } \sin^2 \f }
- {i e^{ 2i \f } \pb \ub  - i e^{-2i \f } \pb u \over
8 \sqrt{1 - u \ub  } \sin^2 \f } \nonumber \\
\bar{a} &=& -i \pb \f - {(e^{ 2i \f } \ub  -u) \pb \ub
-( e^{ -2i \f } u - \ub ) \pb u \over 4 (1-u \ub )}
+{(u-\ub )(\pb u+\pb \ub ) \over 16(1-u \ub )\sin^2 \f}.
\een
Then the flatness condition Eq.(10) resolves into the equations of motion for
$u$, $\ub $ and $\f$. These equations are related to those of the
nonlocal gauge by the following association; if we solve Eq.(10) in terms of
holonomies, $A=h^{-1}\pp h$ and $\bar{A} = h^{-1}\pb h$, then $g_{1}, g_{2}$
of the nonlocal gauge are related to those of the unitary gauge by
\be
g_{1}^{N} = hg_{1}^{U}h^{-1}  \ , \ g_{2}^{N} = hg_{2}^{U}h^{-1} \ .
\ee
For the rest of the Paper, we will restrict ourselves only to the nonlocal
gauge. Translation of subsequent results into the unitary gauge can be
readily made by the association in Eq.(18).

In order to understand the integrability of the matrix sine-Gordon theory,
we consider the linear $4 \times 4$ matrix equations with a spectral parameter
$\l $,
\ben
L_{1}(\l)\Psi &\equiv& ( \pp +  U_0 - \l T )\Psi = 0 \nonumber \\
L_{2}(\l )\Psi &\equiv& (\pb +{\bar {\cal A}}+ \li V_1
)\Psi = 0
\een
where
\be
U_0 =  {\cal G}^{-1}\pp {\cal G}+ {\cal G}^{-1}{\cal A} {\cal G} \ , \
V_{1} = {\cal G}^{-1}\Tb {\cal G} \ , \
T= i\k \S \ , \ \Tb =i\S
\ee
and
\be
{\cal G}  \equiv \pmatrix{
g_1 & 0  \cr 0  & g_2 } \ , \
{\cal A} \equiv \pmatrix{ A &  0 \cr 0  &  A }
\ , \
{\bar {\cal A}} \equiv \pmatrix{ \bar{A} &  0 \cr 0  & \bar{A} }
\ , \
\S \equiv \pmatrix{ 0  & 1 \cr 1 & 0 }
\ee
with each entries being $2 \times 2 $ matrices.
The matrix sine-Gordon equation arises precisely as an integrability
condition, $[L_{1}(\l ) \ , \ L_{2}(\l )]=0$, of the linear equation for any
$\l$. The advantage of the linear equation with a spectral parameter $\l $
is that it allows a systematic way to construct infinite conserved currents.
In addition, exact solutions can be obtained from the linear equation which
we consider in Sec.3. In order to find conserved currents, we solve the linear
equation iteratively by setting
\be
\Phi \equiv \Psi \exp(-\l Tz) = \sum_{m = 0}^{\infty}\l^{-m}\Phi_{m} \ ; \
\Phi_{0} = 1
\ee
so that the $m$-th order equation in the nonlocal gauge is
\ben
\pp \F_{m} &+& U_{0}^{\bot}
\F -[ T, \F_{m+1} ] = 0 \nonumber \\
\pb \F_{m} &+& V_{1}\F_{m-1} = 0 \nonumber \\
U_{0}^{\bot} &\equiv& {1 \over 2} \pmatrix{
\gi_{1}\pp g_{1} - \gi_{2}\pp g_{2}  & 0  \cr
0   & \gi_{2}\pp g_{2} - \gi_{1}\pp g_{1} }   \ .
\een
In accordance with the initial value $\Phi_{0}=1$, $\Phi_{m}$ can be
appropriately parametrized by
\be
\Phi_{2m} \equiv \pmatrix{ P_{2m} &  0  \cr 0 & S_{2m} }
\ , \
\Phi_{2m+1} \equiv \pmatrix{ 0 & P_{2m+1} \cr S_{2m+1} & 0 }
 \ ; \ m \ge 0 \ .
\ee
This brings Eq.(23) into a component form,
\ben
\pb P_{m+1} &+& i\gi_{1}g_{2}S_{m} = 0 \nonumber \\
\pb S_{m+1} &+& i\gi_{2}g_{1}P_{m} = 0 \nonumber \\
\pp P_{m} &+& \gi_{1}\pp g_{1}P_{m} = i\k (P_{m+1}-S_{m+1}) \nonumber \\
\pp S_{m} &+& \gi_{2}\pp g_{2}S_{m} =  i\k (S_{m+1}-P_{m+1})  \ ,
\een
which we solve iteratively with initial values $P_{0} = S_{0} = 1$,
\ben
P_{m+1}-S_{m+1}&=& {1 \over 2i\k }[\pp (P_{m} -S{m})+\gi_{1}\pp g_{1}
(P_{m} +S_{m})] \nonumber \\
P_{m+1}+S_{m+1}&=&-\int d\zb (i\gi_{1}g_{2}S_{m} + i\gi_{2}g_{1}P_{m})
  -\int dz \gi_{1}\pp g_{1}(P_{m+1}-S_{m+1}) \ .
\een
In particular,
\ben
P_{1} &=& {1 \over 2i\k }\gi_{1}\pp g_1 - {1 \over 2i\k }\int dz
(\gi_1 \pp g_1 )^2 - {i \over 2}\int d\bar{z} (\gi_1 g_2 + \gi_2 g_1 )
\nonumber \\
S_{1} &=& -{1 \over 2i\k }\gi_{1}\pp g_1 - {1 \over 2i\k }\int dz (\gi_1
\pp g_1 )^2 - {i \over 2}\int d\bar{z} (\gi_1 g_2 + \gi_2 g_1 ) \ .
\een
With iterative solutions of the linear equation, we find that the consistency
 of Eq.(25), $\pp\pb P_{m} = \pb\pp P_{m}$ and $\pp\pb S_{m} = \pb\pp S_{m}$,
gives rise to two sets of current conservation laws;
\ben
\pb J^{(1)}_{m} &+& \pp \bar{J}^{(1)}_{m+2} = 0   \nonumber \\
\pb J^{(2)}_{m} &+& \pp \bar{J}^{(2)}_{m+2} = 0 \ ; \ \ m \ge 0
\een
where
\ben
\bar{J}^{(1)}_{m} &=& i\gi_{1}g_{2}S_{m}  \nonumber \\
J^{(1)}_{m+2} &= & \pp P_{m+1}
 =  -\gi_{1}\pp g_{1} P_{m+1} + i\k (S_{m+1} - P_{m+1}) \nonumber \\
\bar{J}^{(2)}_{m} &=& i\gi_{2}g_{1}P_{m}  \nonumber \\
J^{(2)}_{m+2} &=& \pp S_{m+1} = \gi_{1}\pp g_{1} S_{m+1} + i\k
(P_{m+1} - S_{m+1}) \ .
\een
The subscript $m$ of the current $J_{m}$ denotes the conformal spin in the
massless limit, or it simply counts the order of the derivatives.
In particular, the $m=0$ case is the energy-momentum conservation,
\be
\pb T_{\pm } + \pp \Theta_{\pm } = 0,
\ee
where
\ben
T_{+} &\equiv & i\k(J^{(1)}_{2} + J^{(2)}_{2}) = (\gi_{1}\pp g_{1})^{2}
\nonumber \\
\Theta_{+} &\equiv & i \k (\bar{J}^{(1)}_{0}+ \bar{J}^{(2)}_{0}) = -\k
(\gi_{1}g_{2} + \gi_{2}g_{1})
\een
while the other half of the conserved currents are
\ben
T_{-} &\equiv & i\k(J^{(1)}_{2} - J^{(2)}_{2}) = \pp(\gi_{1}\pp g_{1})
\nonumber \\
\Theta_{-} &\equiv & i \k (\bar{J}^{(1)}_{0}- \bar{J}^{(2)}_{0}) = -\k
(\gi_{1}g_{2} - \gi_{2}g_{1}) \ .
\een
It is interesting to observe that $T_{-}$ in the abelian limit becomes
$T_{-} = \pp^{2}\phi $ which is precisely the term added to improve the
energy-momentum tensor in the Feigin-Fuchs construction\cite{fuchs}.
Another type of conserved currents arises from the invariance of the matrix
sine-Gordon theory under the parity transform,
\be
z \leftrightarrow \zb  \ \ \mbox{ and } \ \  g_{1} \leftrightarrow \gi_{1} \ ,
\ g_{2} \leftrightarrow \gi_{2} \ .
\ee
This leads to the parity conjugate pair of conserved currents which,
together with currents in Eq.(29), constitute a complete set of
conserved currents of the matrix sine-Gordon theory. For example,
the parity conjugate of the energy-momentum is
\ben
\Tb_{+} &=&  (\pb g_{1}\gi_{1})^{2} \ , \
\bar{\Theta}_{+} = -\k (g_{1}\gi_{2} + g_{2}\gi_{1})
\een
and
\ben
\Tb_{-} = - \pb(\pb g_{1}\gi_{1}) \ , \
\bar{\Theta}_{-} = -\k (g_{1}\gi_{2} - g_{2}\gi_{1})  \ .
\een

\section{Dressing Method and Soliton Solutions}
In this section, we give a detailed account of the derivation of soliton
solutions.
We follow the dressing method of Zakharov and Shabat\cite{zak} and obtain
nontrivial soliton solutions from the trivial one by employing the Riemann
problem technique with zeros\cite{novikov}. In Sec.4, we give an alternative
method based on the B\"{a}cklund transformation and obtain soliton solutions
by direct integration. We first give a brief review on the dressing method.
For later purpose, we rewrite the linear equation in the
nonlocal gauge by making a similarity transform of Eq.(19) by the matrix $Q$,
\be
Q = Q^{-1} \equiv {1 \over \sqrt{2}}\pmatrix{ 1  & 1 \cr
1  & -1 }
\ee
such that
\be
( \pp +  U_0^{'} - \l T^{'} )\Psi^{'} = 0 \ , \
 (\pb + \li V_1^{'}  )\Psi^{'} = 0
\ee
where
\ben
\Psi^{'} &=& Q \Psi Q^{-1} \nonumber \\
U_0^{'} &=& Q U_0 Q^{-1} =\pmatrix{ 0  & g_1^{-1} \pp g_1 \cr
g_1^{-1} \pp g_1   & 0 } \nonumber  \\
V_1^{'}&=& Q V_1 Q^{-1} = {i \over 2} \pmatrix{ g_1^{-1} g_2 + g_2^{-1} g_1
& g_2^{-1} g_1 - g_1^{-1} g_2 \cr
g_1^{-1}g_2 - g_2^{-1} g_1  &  -g_1^{-1} g_2 - g_2^{-1} g_1 }
\een
and
\be
T^{'}= Q T Q^{-1} = {i \kappa} \pmatrix{ 1 & 0 \cr
0  & -1 } \   , \
\Tb^{'}= Q {\bar T} Q^{-1} = {i} \pmatrix{ 1 & 0 \cr
0  & -1 } .
\ee
In the following, we drop the prime for convenience without causing any
confusion.
The dressing method is a systematic way to obtain nontrivial solutions from
a trivial one. In our case, we take the vacuum as a trivial solution of
Eq.(37),
\be
g_1 = g_2 = 1 \  \mbox{ and }
 \Psi = \Psi^{o} \equiv \exp(\l Tz - \l^{-1} \Tb \zb ) \ .
\ee
Let $\Gamma $ be a closed contour or a contour extending to infinity on the
complex plane of the parameter $\l $. Consider the matrix function $\Phi_{+}
(z,\zb ,\l)$ which is analytic with $n$ simple poles $\m_{1}, ... ,\m_{n}$
inside $\Gamma $ and $\Phi_{-}(z,\zb ,\l )$ analytic with $n$ simple
zeros $\l_{1}, ... , \l_{n}$ outside $\Gamma$. We assume that none of these
zeros lies on the contour $\Gamma$ and $\Psi^0 (\Psi^0)^{-1} =
(\Phi_{-})^{-1} \Phi_{+} = 1  $ for $\l \ne \m_{i} ,
\l_{i} \ ; \ i = 1,...,n$.
We normalize $\Phi_{+}, \Phi_{-}$ by  $\Phi_{+}|_{\l = \infty }=
\Phi_{-}|_{\l = \infty } =1$.
Differentiating $ \Psi^{o}\Psi^{o -1} = (\Phi_{-})^{-1}\Phi_{+} = 1 $ with
respect to $z$ and $\zb $, one can easily see that
\ben
\pp \Phi_{+} \Phi_+^{-1} + \l \Phi_{+} T \Phi_{+}^{-1} &=&
\pp \Phi_{-} \Phi_{-}^{-1} + \l \Phi_{-} T \Phi_{-}^{-1}
\nonumber \\
\pb \Phi_{+} \Phi_+^{-1} - \li \Phi_{+} \Tb \Phi_{+}^{-1} &=& \pb \Phi_{-}
\Phi_{-}^{-1} - \li \Phi_{-} \Tb \Phi_{-}^{-1} .
\een
Since $\Phi_{+} (\Phi_{-})$ is analytic inside (outside) $\Gamma$, we find that
the matrix functions $\tilde{U}_0$ and $ \tilde{V}_1$, defined by
\be
\tilde{U}_0 \equiv  -\pp \Phi \Phi^{-1} - \Phi \l T \Phi ^{-1} + \l T
 \ ; \ \tilde{ V}_1  \equiv  -\l \pb \Phi \Phi^{-1} +  \Phi \Tb \Phi^{-1}
\ee
where $\Phi = \Phi_{+}$ or $\Phi_{-}$ depending on the region, become
independent of $\l $. Also, $\tilde{\Psi } \equiv \Phi \Psi^{o}$ satisfies the
linear equation;
\be
( \pp + \tilde{U}_0 - \l T )\tilde{\Psi } = 0 \ , \ (\pb + \li \tilde{V}_1
 )\tilde{\Psi } = 0 \ .
\ee
The identification $\tilde{\J} = \J$ and $\tilde{U}_0 , \ \tilde{V}_1$ with
respect to  $U_0,V_1$ in Eq.(38), $\tilde{U}_0$ and $\tilde{V}_1$ then
provide nontrivial  $n$-soliton solutions.

In making such an identification, the specific form of $U_0$ and $V_{1}$
imposes restrictions on $ \J $ which in certain cases may be solved
algebraically. For example, the anti-unitarity of $U_0$ and $ V_1$ imposes
restrictions on $\J$ and $\F$ which may be complied with
\be
\J^\dagger (\l) = \J^{-1} (\l^* ) \ , \
\F^\dagger (\l) = \F^{-1} (\l^* ) \ .
\ee
These are not the most general expression giving the anti-unitary $U_0$
and $V_1$, however we assume Eq.(44) since they suffice for our purpose of
deriving soliton solutions.
In order to construct the matrix function $\Phi$ for the soliton solutions,
we  take the ans\"{a}tze for $\Phi$ and $\Phi^{-1}$,
\be
\Phi = 1+\sum_{\a =1}^{M} ({A^\a  \over \l -\n_{\a }} ) \ , \
\Phi ^{-1} = 1+\sum_{\a =1}^{M} ({B^\a  \over \l -\m_{\a }} ) ,
\ee
where the matrix functions $A^{\a }(z,\zb ), B^{\a }(z,\zb )$ are to be
determined.
Since the identity; $\Phi\Phi^{-1}=1$ and Eq.(42) should hold for any
$\l $, they require respectively algebraic and differential
relations among  $A^{\a }$ and $B^{\a }$. These relations can be obtained
through the evaluation of residues of both equations at $\l = \m_{\a},
\n_{\a }$. For instance, the residues of the equation
$\Phi\Phi^{-1}=1$ gives rise to
\be
A^{\a } + A^{\a } \sum_{\b=1}^{M}({B^{\b} \over \n_{\a }-\m_{\b}})
			      = 0 \ , \
B^{\a }+\sum_{\b=1}^{M}({A^{\b} \over \m_{\a }-\n_{\b}})B^{\a } = 0 ,
\ee
while those of Eq.(42) lead to
\be
A^{\a } D_{1,2} (\n_{\a }) [1 + \sum_{\b =1}^{M}({B^{\b } \over \n_{\a }
-\m_{\b}})] = 0 \ , \
[1 + \sum_{\b =1}^{M}({A^{\b } \over \m_{\a }-\n_{\b }})] D_{1,2} (\m_{\a})
B^{\a } = 0
\ee
where
\be
D_1(\l ) \equiv \pp - \l T
 \ ,  \
D_2(\l ) \equiv \pb + \li \Tb .
\ee
In order to solve Eqs.(46) and (47), we assume that
$A^{\a}_{ij} = s^{i}_{\a}t^{j}_{\a} \ , \ B^{\a}_{ij} = n^{i}_{\a}m^{j}_{\a}$
where $s^{i}_{\a}, t^{i}_{\a} ,  n^{i}_{\a}, m^{i}_{\a}$ are two by two
matrices with $i=1,2$.
Then, Eqs.(46) and (47) changes into
\be
\sum_{j=1}^{2} t_{\a}^{j} [\d^{ji} + \sum_{\b=1}^{M}{n_{\b}^{j}m_{\b}^{i}
\over \n_{\a}-\m_{\b} } ]  = 0 \ , \
\sum_{i=1}^{2} [\d^{ji} + \sum_{\b=1}^{M}{s_{\b}^{j}t_{\b}^{i}
\over \m_{s}-\n_{\b} } ] n_{s}^{i}  = 0
\ee
and
\be
D_{1,2}(\m_{\a}) n_{\b} =0 \ , \ t_\b D_{1,2}(\n_{\a}) = 0  \ ,
\ee
where we understand $t_\b \pp $ and $t_\b \pb $  as $-\pp t_\b $ and $-\pb
t_\b$.
Note that  $n_{\a}$ and $t_{\a}$ can be solved in terms of arbitrary constant
vectors $\bar{n}_{\a}$ and $\bar{t}_{\a}$,
 \be
n_{\a}^{i} = \sum_{j=1}^{2}[\J^{o} (\m_\a) ]^{ij} \bar{n}^{j}_{\a} \ , \
t^{i}_{\a} =\sum_{j=1}^{2} [ \J^{o}(\n_{\a})^{-1} ]^{ji} \bar{t}^{j}_{\a} \ ,
\ee
while  $m_{\a}$ and $s_{\a}$ can be  obtained in terms of $t_{\a}$ and $n_{\a}$
by
solving the linear algebraic equation (49) such that
\be
m^{i}_{\a} = - \sum_{\b=1}^{M} V^{-1}_{\a \b} t^i_\b \ , \
s^{i}_{\a} =  \sum_{\b=1}^{M} n^i_\b V^{-1}_{\b \a} \ ,
\ee
where
\be
V_{\a \b} \equiv \sum_{j=1}^{2} {t_{\a}^{j} n_{\b}^{j} \over
\n_{\a} - \m_{\b} }  \ .
\ee
$V^{-1}$ is defined by $\sum_{\b=1}^{M} V^{-1}_{\a \b} V_{\b \g} =
\d_{\a \g} 1$ where $1$ is the $2\times 2$ unit matrix.
The unitarity condition Eq.(44) requires that
\be
\m_{\a} = \n^{*}_{\a} \ \  \mbox{ and } \ \  A^{\a \dagger } = B^{\a} (
\ \ \mbox{ or } \ \  t^{i \dagger}_{\a} = n^{i}_{\a}, \  s^{i \dagger}_{\a}
= m^{i}_{\a} \ ).
\ee
Consequently,
\be
\bar{t}^{i \dagger}_{\a} = \bar{n}^{i}_{\a} \ , \
(V^{-1}_{\b \a} )^\dagger = - V^{-1}_{\a \b} \ .
\ee
Further specification of $t_{\a}$ and $n_{\a }$ arises from the identification;
$U_{0}=\tilde{U}_{0} , V_{1}=\tilde{V}_{1}$, in Eq.(42).  Since Eq.(42) holds
for any $\l $, we combine Eq.(42) and (45) and take the
$\l \rightarrow  \infty$ limit to obtain
\be
-U_0 = \sum_{\a } (A^{\a } T + T B^{\a })
= i \k \sum_{\a} \pmatrix{ s^{1}_{\a} t^{1}_{\a} + n^{1}_{\a} m^{1}_{\a}
& -s^{1}_{\a} t^{2}_{\a} + n^{1}_{\a} m^{2}_{\a} \cr
s^{2}_{\a} t^{1}_{\a} - n^{2}_{\a} m^{1}_{\a}
& -s^{2}_{\a} t^{2}_{\a} - n^{2}_{\a} m^{2}_{\a} }
\ .
\ee
Note that the (block)-diagonal part vanishes identically  due to the equality
\be
s^{i}_{\a} t^{i}_{\a} + n^{i}_{\a} m^{i}_{\a} =
n^{i}_{\b} (V^{-1})_{\b \a} t^{i}_{\a} - n^{i}_{\a} (V^{-1})_{\a \b}
t^{i}_{\b} =0
\ee
which agrees with $U_{0}$. The off-diagonal part gives rise to
\be
\gi_{1}\pp g_{1}   = -i\k ( -s^{1}_{\a} t^{2}_{\a} + n^{1}_{\a} m^{2}_{\a} ) =
2i\k \sum_{\a , \b} n^{1}_{\b} V^{-1}_{\b \a} t^{2}_{\a}  =
 2 i \k \sum_{\a , \b}  Z^{-1}_{\a \b} ,
\ee
or
\be
g_1^{-1} \pp g_1 = - i\k(s^{2}_{\a} t^{1}_{\a} - n^{2}_{\a} m^{1}_{\a} )
= -2i\k \sum_{\a , \b } n^{2}_{\a} V^{-1}_{\a \b} t^{1}_{\b} =
-2 i \k \sum_{\a , \b}{\tilde Z}^{-1}_{\a \b} \ ,
\ee
where $Z_{\a \b}$ and ${\tilde Z}_{\a \b}$ are defined by
\ben
Z_{\a \b} &\equiv & { (t^2_\b)^{-1} t^1_\b + n^2_\a (n^1_\a)^{-1} \over
{\n_\b - \m_\a} } =  { N^\dagger_\b + N^{-1}_\a \over \m^{*}_\b - \m_\a }
\nonumber \\
{\tilde Z}_{\a \b} &\equiv& { n^1_\a (n^2_\a)^{-1} +(t^1_\b)^{-1} t^2_\b
\over {\n_\b - \m_\a} } =  { N_\a + N^{-1 \dagger}_\b
\over \m^{*}_\b - \m_\a } = -Z^{\dagger}_{\b\a } \nonumber \\
N_\b &\equiv& n^1_\b (n^2_\b)^{-1} =
\exp(2 \D_\b) \bar{n}^1_\b (\bar{n}^2_\b)^{-1} \equiv \exp(2 \D_\b)
\bar{N}_\b \nonumber \\
\D_\b &\equiv& i \k \m_\b z - { i\zb \over \m_\b } \
\een
and we have used the unitarity condition Eq.(54). The last step in Eq.(58)
(similarly Eq.(59)) can be checked easily by using
$K^{2}_{\a} \equiv \sum_{\b} V^{-1}_{\a \b} t^{2}_{\b}$
so that
\be
t^2_{\b} = \sum_{\a} V_{\b \a} K^{2}_{\a} =
t^{2}_{\b} \sum_{\a} { (t^{2}_{\b})^{-1} t^{1}_{\b} +
n^{2}_{\a} (n^{1}_{\a})^{-1} \over \n_{\b} - \m_{\a} } n^{1}_{\a} K^2_{\a} =
t^{2}_{\b}
\sum_{\a} Z_{\a \b }n^{1}_{\a}K^2_{\a} \ .
\ee
Thus, the identification with $U_0$ through Eqs.(58) and (59) imposes
restrictions
on $Z_{\a \b }$ such that
\be
\sum_{\a , \b} Z^{-1}_{\a \b} = - \sum_{\a , \b} {\tilde Z}^{-1}_{\a \b}
=\sum_{\a , \b} Z^{\dagger -1}_{\a \b} .
\ee

On the other hand, the $V_1$ part in the $\l \rightarrow 0$ limit
of the linear equation, gives rise to
\be
V_1 = \F \bar{T} \F^{-1}|_{\l = 0 } \  \mbox{  or } \  \F |_{\l = 0 }
\bar{T} = V_1 \F |_{\l = 0 } \ .
\ee
In components, they are
\be
g_1^{-1} g_2 = -(\F_{12} + \F_{22} ) ( \F_{12} - \F_{22} )^{-1}
\ee
or
\be
g_1^{-1} g_2  = - ( \F_{11} + \F_{21} ) ( \F_{21} - \F_{11} )^{-1}
\ee
where $\F_{ij}$ denote $2\times 2$ block components of $\F |_{\l = 0 }$.
We will see below that the two expressions for $g_1^{-1} g_2$ in Eqs.(64) and
(65)
are indeed equivalent when the condition Eq.(62) is satisfied. If we define
\be
Y_{\a \b} \equiv { 1+N^\dagger_\a N_\b \over 1- \m_\b/\m^{*}_\a } \ ,
\ee
Eqs.(64) and (65) become
\ben
g_1^{-1} g_2 &=& -(\F_{12} + \F_{22} ) ( \F_{12} - \F_{22} )^{-1} \nonumber \\
&=& \biggl[ 1- \sum_\a { (s^1_\a + s^2_\a) t^2_\a \over \n_\a} \biggr]
  \biggl[ 1- \sum_\a { (-s^1_\a + s^2_\a) t^2_\a \over \n_\a} \biggr]^{-1}
\nonumber \\
&=& \biggl[ 1-\sum_{\a,\b}(1+N_\b)(Y^{-1})_{\b \a} \biggr]
 \biggl[ 1-\sum_{\a,\b}(1-N_\b)(Y^{-1})_{\b \a}\biggr]^{-1},
\een
and
\ben
g_1^{-1} g_2 &=& - ( \F_{11} + \F_{21} ) ( \F_{21} - \F_{11} )^{-1} \nonumber
\\
&=& \biggl[ 1- \sum_\a { (s^1_\a + s^2_\a) t^1_\a \over \n_\a} \biggr]
  \biggl[ 1 +\sum_\a { (-s^1_\a + s^2_\a) t^1_\a \over \n_\a} \biggr]^{-1}
\nonumber \\
&=&\biggl[ 1-\sum_{\a,\b} (1+N_\b ) (Y^{-1})_{\b \a} N^\dagger_\a ]
[1 + \sum_{\a,\b}(1-N_\b )(Y^{-1})_{\b \a} N^\dagger_\a \biggr]^{-1} \ .
\een
These are M-soliton solutions of the matrix sine-Gordon theory. In the
following, we give an explicit expression for M=1 and 2.
\vskip .5in
\underline{ \bf M=1;  1-soliton}
\newline
\vskip .3in
For M=1, we have
\be
Z= -\tilde{Z}^{\dagger} = { N^\dagger + N^{-1} \over \m^* - \m } \ee
and Eq.(62) for M=1, $Z^{-1} = - \tilde{Z}^{-1}$, can be solved either by
$N=-N^{-1}$ or $N= -N^\dagger$. The former case results in only a trivial
solution while the latter case, $N = \exp(2 \D) \bar{N} =
-N^{\dagger}=-\exp(2 \D^*) \bar{N}^\dagger$, requires $\D$ real and
$\bar{N}$ anti-hermitian. Thus, $\m \equiv i \d$ is pure imaginary
and $\D = -\k \d z - \zb /\d $. We parameterize the anti-hermitian matrix
$\bar{N}$ by $\bar{N}= i \exp(\h ) a_i \s_i$ where $\s_{i}$
are Pauli matrices and the repeated index $i$ denotes summation from i=1 to 3.
$\h , a_{i}$ are arbitrary real constants with a normalization
$ a_i a_i = 1$.  Then, from Eqs.(67) and (58) we
obtain the 1-soliton solution given by
\be
g^{-1}_1 g_2 = \left( { 1+ N \over 1 - N } \right) ^2
= \left( - \tanh (2\D + \h) + i a_i \s_i
{1 \over \cosh (2 \D + \h)} \right)^2 .
\ee
and
\be
g^{-1}_1 \pp g_1 = 2 i \k Z^{-1} =
2 i \k \d { 1 \over \cosh (2 \D + \h) } a_i \s_i \ .
\ee
Combining Eqs.(70) and (71), we could solve for $g_{1}$ and $g_{2}$. They agree
with the explicit form given in Eq.(131) of Sec.4 which is derived directly
from
the B\"{a}cklund transform.
Physical meaning of parameters in the soliton solution is the following;
parameter $\h$  depends on the choice of origin of space and time.
We choose the origin to set them to zero and introduce the space and time
coordinate by $t \equiv z + \bar{z},\ \ x\equiv  z - \bar{z} $.
Parameter $\d$ describes the velocity $u$ of the soliton where
$u = ( 1 +  \k \d^2 )/ ( 1 - \k \d^2 )$. Then,
\be
\D =\pm {  \sqrt{-\k}  \over \sqrt{1 - u^2}}(x - ut) \
\ee
where $\pm $ denotes the sign of $\d $.

A few remarks are in order.

\noindent
(i) For $N= -N^\dagger$, the two expressions of $\gi_{1}g_{2}$
given in Eqs.(67) and (68) yield the same result, Eq.(70), thus proving the
consistency
of two expressions in this case.

\noindent
(ii) We may obtain an abelian limit by taking
$a_{1}=a_{2}=0, \  a_{3}=1$ and $g_{1}=\gi_{2} =\exp(i\sqrt{\pi } \varphi
\s_{3}) $.
In which case, Eq.(70) reduces to the well-known 1-soliton solution of the
sine-Gordon equation\cite{raj},
\be
\varphi = -{2 \over \sqrt{\pi }}\tan^{-1} e^{2\D  }   \ .
\ee

\noindent
(iii) In the parametrization  $\gi_{1}g_{2} =  \exp(i\f a_i\s_i )$, 1-soliton
can be written by
\be
\f = 2\cos^{-1}(-\tanh{2\D  }) = 2\sin^{-1} {1 \over \cosh {2\D }} \ .
\ee
Note that $\D$ changes from $\pm \infty $ to $ \mp \infty $  as $x$ goes from
$-\infty $ to  $\infty $ so that the soliton number,  $(\f ({\infty })
-\f (-\infty ))/2\pi$, is $ \pm 1$.
\vskip .5in
\underline{\bf M=2; soliton(antisoliton) - soliton(antisoliton) scattering }
\vskip .3in
For M=2, two possible solutions of Eq.(62) are
\ben
\mbox{(i) } \ \ N^\dagger_1 &=& - N_1,\ N^\dagger_2 = -N_2  \nonumber \\
\mbox{(ii)} \ \ N_{1}^{\dagger }&=& -N_{2} ,
\een
which describe 2-soliton solutions and
nonabelian breather solutions respectively. First, we consider the case (i).
We parametrize $N_{1}, N_{2}$ by
\be
N_1 = i \exp(2 \D_1 + \h_1) a_i \s_i \ , \
N_2 = i \exp(2 \D_2 + \h_2) b_i \s_i
\ee
where $\m_{k} = i\d_{k}, \   \D_{k} = -\k \d_{k} z - \zb /\d_{k} $ and
$a_i, b_i,\ \h_i$ are real constants with  normalzation
$a_i a_i = b_i b_i = 1$. In order to check
that the criterion (i) indeed satisfies Eq.(62), we note that,
for example, $Z_{21} Z^{-1}_{22} Z_{12} = Z_{12} Z^{-1}_{22} Z_{21}$ due to
the property that $N_i N^\dagger_i = \exp(4 \D_i + 2 \h_i)$ which is
proportional
to the identity matrix. Therefore,
\be
(Z^{-1})^\dagger_{11} = (Z^\dagger_{11} - Z^\dagger_{21} Z^{-1 \dagger}_{22}
Z^\dagger_{12})^{-1} = (Z_{11} - Z_{21} Z^{-1}_{22} Z_{12})^{-1}  =
(Z_{11} - Z_{12} Z^{-1}_{22} Z_{21})^{-1}= (Z^{-1})_{11}  \ .
\ee
Similar procedure for other components of $Z^{-1}$ leads to Eq.(62).

We now calculate $g^{-1}_1 g_2$ for the 2-soliton solution.
{}From Eqs.(66) and (76), we have
\be
Y^{-1} =
(\det Y)^{-1} \pmatrix{ {1\over 2}( 1 -N_{2}^{2} ) &
(N_1 N_2 -1 )( 1 + {\d_2 \over \d_1 })^{-1}  \cr
(N_2 N_1 -1 )( 1 + {\d_1 \over \d_2 })^{-1}  &
{1\over 2}( 1 - N_{1}^{2}) }
\ee
where
\be
\det Y = {1\over 4} ( 1 + e^{4 \D_1 + 2 \h_1}) (1 + e^{4 \D_2 + 2 \h_2})
-{\d_1 \d_2 \over  (\d_1 +\d_2 )^{2}} (1 - N_1 N_2) (1 - N_2 N_1) \ .
\ee
Thus, $g^{-1}_1 g_2$ can be readily calculated from Eq.(67) or Eq.(68).
Either case gives rise to the same result in the form;
\ben
g^{-1}_1 g_2 &=& ( A + B^i_{+} \s_i ) (A + B^i_{-} \s_i )^{-1} \nonumber \\
&\equiv & M_0 + M_i \s_i  \ .
\een
With the notation,
\be
\D_{\pm} \equiv (\D_{1} + {1\over 2}\h_{1}) \pm (\D_{2} +  {1\over 2}\h_{2}) \
, \
R \equiv { \d_1-\d_2 \over  \d_1+\d_2 } \ ,
\ee
each coefficients are given by
\ben
A &\equiv& e^{2\D_{+}}
[ R^{2}\cosh^{2}\D_{+} -\sinh^{2}\D_{-} +
{1+R^2 \over 2}(a_i b_i -1)] \\
B^i_{\pm} &\equiv& iRe^{2\D_{+}}[\mp a_{i}\sinh (2\D_{2}+\h_{2}) \pm
b_{i}\sinh (2\D_{1}+\h_{1}) +  \e_{ijk} a_j b_k  ]
\een
and
\ben
M_0 &=& { A^2 - B^k_{+} B^k_{-} \over A^2 - B^k_{-} B^k_{-} } \nonumber \\
    &=& 1 - { 4 R^{2}\over P_{+}^{2}} \biggl[ 2 \sinh^2 \D_{-} \cosh^2 \D_{+} +
	(\cosh^2 \D_{+} -\cosh^2 \D_{-})(1-a_i b_i) \biggr] \\
M_i &=& { A (B^i_{+} - B^i_{-}) - i\e_{ijk} B^j_{+} B^k_{-}
\over A^2 - B^k_{-} B^k_{-} } \nonumber \\
   &=&  {2iR \over P_{+}^{2}} \biggl[ \{ Rb_{i}-(P_{-} + Ra_k b_k )a_{i}\}
   \sinh (2\D_{2}+\h_{2})
   + \{ Ra_{i} + (P_{-}-Ra_{k}b_{k})b_{i} \}
   \sinh (2\D_{1} +\h_{1})\biggr]  \nonumber \\
\een
where
\be
P_{\pm } \equiv  (R^{2}\cosh^{2}\D_{+} \pm \sinh^{2}\D_{-} \mp
{1\mp R^2 \over 2}(a_i b_i -1))
\ee
As in the 1-soliton case, we make the choice of the origin of the coordinate
to set parameters $\h_i$ to zero. Parameters $\d_i$ also describe the velocity
of solitons. As we show below, if $\d_2 = 1 / \k \d_{1} $, it descibes the
soliton - soliton,  or antisoliton -
antisoliton scattering, whereas if $\d_2 = -1 / \k \d_{1} $, it descibes
the soliton - antisoliton scattering in the center of mass frame. In both
cases, velocities of each solitons are given by
$u = ( 1 +  \k \d_{1}^2 )/ ( 1 - \k \d_{1}^2 )$ and $-u$. In the soliton -
soliton scattering case, $R = -1/u \ , \
\sqrt{-\k }\d_1 = \mp \sqrt{1-u}/\sqrt{1+u}$ and
\ben
\D_{-}&=& \D_1 - \D_2 = \mp { 2 \sqrt{-\k} x \over \sqrt{1 - u^2}} \equiv
\mp X \nonumber  \\
\D_{+}&=& \D_1 + \D_2 = \pm { 2 \sqrt{-\k}  ut \over \sqrt{1 - u^2}} \equiv \pm
T \ .
\een
The  upper sign corresponds to the soliton - soliton and
the lower sign to the antisoliton - antisoliton scatterings respectively.
In the soliton - antisoliton case, $R = -u$ and
\ben
\D_{-}&=& \D_1 - \D_2 = \pm { 2 \sqrt{-\k} ut \over \sqrt{1 - u^2}}=
\pm T \nonumber \\
\D_{+}&=& \D_1 + \D_2 = \mp { 2 \sqrt{-\k} x \over \sqrt{1 - u^2}} = \mp X \ .
\een
where the upper(lower) sign corresponds to the soliton(antisoliton) -
antisoliton(soliton) scattering.
In order to convince the correctness of the solution given by Eqs.(80) - (85),
we have checked explicitly that Eqs.(80) - (85), together with
\ben
g^{-1}_1 \pp g_1 &=& 2 i \k \sum_{\a,\b} (Z^{-1})_{\a \b} \nonumber \\
&=& 2 \k \{ (Y^{-1})_{11} \d_1 N_1 + (Y^{-1})_{12} \d_2 N_2 + (Y^{-1})_{21}
\d_1 N_1
 + (Y^{-1})_{22} \d_2 N_2 \} \ ,
\een
indeed satisfy the matrix sine-Gordon equation (12). However, instead of giving
 cumbersome details of the calculation, we present another consistency
check. In the abelian limit, where we take
$a_1 = b_1 = 1, \ \ a_2 = a_3 = b_2 = b_3 = 0$ and $g_1 = \gi_2 =
\exp( i \sqrt{\pi} \varphi \s_1)$, Eqs.(80) -(85) gives rise to
\be
\tan (2 \sqrt{\pi} \varphi ) = { i M_1 \over M_0 } = { iA ( B^1_{+} - B^1_{-} )
\over A^2 - B^1_{+} B^1_{-} } = {2 i A B^1_{+} \over A^2 +(B^1_{+})^2 }
\ee
and
\be
B^1_{+} = \pm { 2i  \over u }\exp (\pm 2 T) \sinh X \cosh T \ , \
A =\exp (\pm 2 T)( {1 \over u^2} \cosh^{2} T -  \sinh^{2}X ) \ .
\ee
Using the identity,
\be
\tan 4 \q = { \sin 4 \q \over \cos 4 \q} =
{4 \sin \q \cos \q ( \cos^2 \q - \sin^2 \q ) \over ( \cos^2 \q -
\sin^2 \q )^2 - 4 \sin^2 \q \cos^2
\q} ,
\ee
we obtain
\be
\varphi = { 1 \over 2 \sqrt{\pi} }\tan^{-1} (- \tan 4 \q) =  { 1 \over 2
\sqrt{\pi} } (-4 \q)
 = \pm { 2 \over  \sqrt{\pi} } \tan^{-1} {u \sinh X \over \cosh T }
\ee
which is precisely the 2-soliton solution of the sine-Gordon theory
for the soliton-soliton scattering with the plus sign and the
antisoliton-antisoliton scattering with the minus sign\cite{raj}.

In order to have a pictorial description of scattering of solitons,
we take without loss of generality,
$\{ a_i \} = (1,0,0)\ ,  \ \{ b_i  \} = ( \cos \q, \sin \q, 0)$. Then,
\ben
M_0 &=& 1 - { 4 \over u^2 P_{+}^{2}} \biggl[ 2 \sinh^2 X \cosh^2 T +
(1 - \cos \q)(\cosh^2 T -\cosh^2 X)\biggr]
\nonumber \\
M_1 &=&\pm {2i \over uP_{+}^{2}} \biggl[ P_{-}\{  \sinh (T+X) - \cos \q \sinh
(T-X) \}
+ {1 \over u }(1 - \cos^2 \q) \sinh (T-X) \biggr] \nonumber \\
M_2 &=& \pm {2i \over u^{2}P_{+}^{2}} \biggl[ - uP_{-} \sin \q  \sinh (T-X)
 + \sin \q \sinh(T+X) -  \cos \q \sin \q \sinh (T-X)\biggr]   \nonumber \\
M_3 &=& 0
\een
where
\be
P_{\pm } \equiv {1 \over u^2} \cosh^2 T \pm \sinh^2 X \pm
{ 1 \over 2} ( 1 - \cos \q)( 1 \mp {1 \over u^2} ) \ .
\ee
The internal motion of solitons may be described most naturally in terms of the
parametrization $g^{-1}_1 g_2 \equiv \exp(-2 i \sqrt{\pi} \varphi_i \s_i )$
where
$\varphi_{i}$ is related to $M_{i}$ by
\be
M_{0}=\cos 2\sqrt{\pi} |\varphi | \ , \ M_{i} = -i {\varphi_{i} \over
|\varphi |}\sin 2\sqrt{\pi } |\varphi | \  ; \ i = 1,2  \ .
\ee
Figures (1)-(3) show $M_{0} \ , \ M_{1}/i\  , \  M_{2}/i $, as an example, for
a specific case where
$\theta = 0.4 \pi $ and the velocity $u = 0.1$. From Eq.(7), the potential
energy
can be written by $V = 2N\k M_{0}/\pi$ so that $M_{0}$
depicts the trajectory of soliton - soliton scattering in terms of minus the
potential energy. $M_{0}$ shows that two solitons repulse each other at the
origin. It is easy to read the $2\pi $-angle variation $\D |2\sqrt{\pi }\varphi
|=2\pi $
across each bump  from these figures which shows clearly that they describe the
soliton - soliton scattering. Note that the internal direction given by the
vector $\{ \varphi_{i} \}$ changes after the collision. Under the
spacetime inversion; $(T, X) \leftrightarrow (-T,-X)$, the solution in Eq.(94)
possesses symmetry; $M_{0} \leftrightarrow M_{0} , M_{i} \leftrightarrow
-M_{i} \ ; \ i=1,2 $. Thus, the internal directions of each solitons, specified
by
the components $\varphi_{i}$, become exchanged in the process of scattering.
This is a characteristic of the scattering of nonabelian solitons.
The minimum points of $M_0$ constitute a trajectory of the center of each
solitons. At time $T= -T_{0}$ and $T=T_{0}$, two solitons are located at
$\pm X_{0}$ which satisfies the relation,
\be
\sinh{X_{0}^{2}}
= { 2 ( \cos \q -1) ( \cosh^2 T_{0} -1) \over 2 \cosh^2 T_{0} -1 + \cos \q }
+ { 1 \over u^2} \cosh^2 T_{0} + { 1 \over 2} ( 1 - \cos \q) ( 1 - { 1
\over u^2}).
\ee
Notice that at $T_{0}=0$, two solitons approach closest with
\be
\sinh{X_{0}^{2}} = {1 \over u^2}
{ 1 + \cos \q \over 2} + { 1 - \cos \q \over 2} \ .
\ee
This shows that the repulsion between two solitons becomes maximum when two
vectors $a_i,\ \ b_i$ are aligned in the same direction which is precisely
the abelian case where $\q = 0$. If $\q = \pi$, $X_0$ takes a minimum value
thereby maximizing the nonabelian effect. On the other hand, when $T_0$ becomes
 large, Eq.(97) can be approximated by
\be
\sinh X_{0}^{2} \approx  { 1 \over u^2} \cosh^2 T_0   \ .
\ee
The elapsing time for two solitons to bounce back to the separating
distance $2X_{0}$ is given by $2T_{0}$. Thus, when $X_{0}$ becomes large,
the elapsing time becomes independent of the angle $\q $.
\vskip .5in
\underline{\bf M=2, nonabelian breather}
\vskip .3in
Now we consider the case (ii); $N^\dagger_1 = -N_2$ in Eq.(75).
We take $\m_1 = - \m^*_2 \equiv { i \over \sqrt{-\k} } \exp (i \a)  $
and parameterize $N_{1}$ and $N_{2}$ by
\ben
N_1 &=& i \exp (2 \D ) (c_k + i d_k ) \s _k \nonumber \\
N_2 &=& i \exp (2 \D ^*  ) ( c _k - i d _k) \s _k
\een
where $c_i c_i = d_i d_i =1$\footnote{This normalization merely dictates the
choice of the
origin of coordinates $x$ and $t$.} and
\ben
2 \D  &=& 2 \sqrt{- \k}[ \exp (i \a) z - \exp (-i \a) \bar{z}]
= 2 \sqrt{- \k} [ (\cos \a) x + i (\sin \a) t ] \nonumber \\
&\equiv& K (x+ivt) \ \ ; \ \ K \equiv { 2 \sqrt{- \k } \over
\sqrt{1 + v^2} } \ , \ v \equiv \tan \a  \ .
\een
A straightforward calculation shows that Eq.(62) holds for
$N_{1}, N_{2}$ given in Eq.(100). We could follow a similar procedure as in
the case of the two soliton scattering and make use of the  fact;
$Z ^ \dagger _{12} = Z _{21}, \ \ Z ^{\dagger} _{11} = Z _{22}$
and $N ^{2} _{1}$ is proportional to the identity matrix.

{}From Eqs.(66) and (100), we have,
\be
Y^{-1} =
(\det Y)^{-1} \pmatrix{ (1 + N_1 N^\dagger _1 )[ 1 + \exp (-2 i \a ) ]^{-1}
&  i c_i d_i \exp( 4 \D^* ) -{1 \over 2 }  \cr
-i c_i d_i \exp(4 \D)-{1 \over 2 }
& ( 1 + N^\dagger_1 N_1 )[ 1 + \exp( 2 i \a ) ]^{-1} }
\ee
where
\be
\det Y = { v^2 \over 4} [1+ 4 (c_i d_i)^2 e^{4Kx}]+ (1 + v^2) e^{2 Kx}
+c_i d_i e^{2Kx} \sin(2Kvt).
\ee
Then, $\gi_{1}g_{2}$ for the breather solution can be obtained from
Eq.(67), or consistently from Eq.(68), which we write in the form;
\ben
g^{-1}_1 g_2 &=& ( A + B^i_{+} \s_i ) (A + B^i_{-} \s_i)^{-1} \nonumber \\
&\equiv& M_0 + M_i \s_i
\een
where
\ben
A &=& { v^2 \over 4} [ 1 + 4 (c_i d_i)^2 e^{ 4 K x } ] + (v^2-1)
e^{2 K x} -c_i d_i e^{2 K x} \sin 2 K v t \\
B^i_{\pm} &=& \pm i c_i v ( e^{Kx} \sin Kvt + 2 c_k d_k e^{3Kx} \cos Kvt)
-2 i v e^{2Kx} \e_{ijk} c_j d_k.
\nonumber \\
&& \pm  i d_i v ( e^{Kx} \cos Kvt + 2 c_k d_k e^{3Kx} \sin Kvt)
\een
and
\ben
M_0 &=& 1 - {2 v^2 \over (\det Y)^2 } \biggl[ e^{2Kx} \{ 1+4 (c_i d_i)^2
e^{4Kx}\}
(1 + c_i d_i \sin 2Kvt)
+ 4 c_i d_i e^{4Kx} ( c_i d_i+\sin 2Kvt)\biggr]  \nonumber \\
M_i &=& { 1 \over (\det Y)^2 }\biggl[ (e^{Kx} \sin Kvt + 2 c_k d_k e^{3Kx}
\cos Kvt ) \{ 2 i A c_i v + 4 i v^2 e^{2Kx} (d_i - c_k d_k c_i)\}
\nonumber \\
&&+ (e^{Kx} \cos Kvt + 2 c_k d_k e^{3Kx} \sin Kvt)
\{ 2iA d_i v + 4 i v^2 e^{2Kx} (c_k d_k d_i-c_i)\} \biggr] .
\een
In addition, a straightforward calculation shows that
\ben
g^{-1}_1 \pp g_1 &=& 2 i \k \sum_{\a , \b} (Z^{-1})_{\a \b} \nonumber \\
&=& - {2i \sqrt{- \k} \over \det Y } \biggl[ i e^{i \a + 2\D   } (- {v \over 2}
+
i v c_k d_k e^{4 \D^* } ) ( c_j + i d_j ) \s_j  + c.c. \biggr]  \ ,
\een
which together with Eqs.(104)-(107) satisfies the matrix sine-Gordon equation
(12).

For a pictorial description of a nonabelian breather, we choose without loss
of generality $\{ c_i \} = (1,0,0)\ ,  \ \{ d_i  \} = ( \cos \q, \sin \q, 0)$.
Then,
\ben
M_0 &=& 1 - {2 v^2 \over P_{+}^2 }\biggl[ e^{2Kx} ( 1+4 \cos^{2}\q e^{4Kx})
(1 + \cos \q \sin 2Kvt)
+ 4 \cos \q e^{4Kx} ( \cos \q + \sin 2Kvt)\biggr] \nonumber \\
M_1 &=& { 2iv \over P_{+}^2 } \biggl[ (P_{-}e^{Kx} + 2\cos^{2} \q P_{-} e^{3Kx}
- 4v \cos \q \sin^{2} \q e^{5Kx} ) \sin Kvt \nonumber \\
&& + (\cos \q P_{-} e^{Kx} + 2\cos \q P_{-} e^{3Kx} - 2v \sin^{2}\q e^{3Kx})
\cos Kvt \biggr]
\nonumber \\
M_2 &=& { 2iv \over P_{+}^2 } \biggl[ (\sin 2\q P_{-}e^{3Kx} + 2v\sin \q
e^{3Kx}
+2v \cos \q \sin 2 \q e^{5Kx} ) \sin Kvt \nonumber \\
&& + (P_{-}\sin \q e^{Kx} + v\sin 2\q e^{3Kx} + 2v\sin 2\q e^{5Kx} )
\cos Kvt \biggr]  \nonumber \\
M_{3} &=& 0 \ ,
\een
where
\be
P_{\pm } = { v^2 \over 4} [1+ 4 \cos^{2}\q  e^{4Kx}]\pm
(1 \pm v^2) e^{2 Kx} \pm \cos \q  e^{2Kx} \sin(2Kvt).
\ee
Figures (4)-(6) show $M_{0} \ , \  M_{1}/i \ , \ M_{2}/i $ for a specific case
where
$\q =  \pi /6 $ and $v = 0.1$.
The potential energy profile given in terms of $M_{0}$ shows clearly the
breathing motion. The behavior of $\varphi_{i}$ in Eq.(96) along the
$X$-direction in the figures of $M_{0}, M_{1}$ and $M_{2}$ confirms that the
breather solution is indeed a bound
state of soliton and antisoliton. In addition, $M_{1}$ and $M_{2}$ shows that
the internal direction of the breather also oscillates which is a
characteristic
 of a nonabelian breather. Two particular values of $\q $ are worth to address.
If $\q = 0$, we may follow a similar procedure as in the 2-soliton case,
 and see that the nonabelian breather reduces to the well known
sine-Gordon breather,
\be
\varphi  = \pm { 2 \over  \sqrt{\pi} } \tan^{-1} {\sin(Kvt + {\pi \over 4})
\over v\cosh(Kx + \mbox{ln}\sqrt{2}) } \ .
\ee
For $\q = \pi /2$, it is interesting to note that $M_{0}$ becomes independent
of $t$ while $M_{1}$ and $M_{2}$ are not. This shows that the nonabelian
breather at $\q = \pi/2 $ breathes only internally. That is, the internal
direction
oscillates while the potential energy remains static, i.e. externally it
becomes
completely breatheless.

\section{B\"{a}cklund Transformation}
The B\"{a}cklund transformation(BT) is a mapping between two solution surfaces
of
certain differential  equations.  For example, the sine-Gordon equation
$\pp\pb \f = \sin \f $ is invariant under the BT
\ben
\pp \tilde{\f} &=& \pp \f - 2\d \sin({\f + \tilde{\f} \over 2 }) \nonumber \\
\pb \tilde{\f} &=& - \pb \f + {2\over \d } \sin({\f -\tilde{\f} \over 2 })
\een
where $\d $ is a nonzero real parameter. The integrability of
Eq.(112) is the requirement that $\f $ and $\tilde{\f} $ are both solutions of
the sine-Gordon equation. Thus
the BT generates a new solution from a known one. Moreover, through the
Bianchi's
permutability theorem, it leads to a nonlinear superposition of solutions
which gives rise to a new solution by purely algebraic means. For example,
if $\f_{a}, \f_{b}$ are two solutions generated by the BT from a known solution
$\f_{g}$ with B\"{a}cklund parameters $\d_{1} , \d_{2}$ respectively, then the
Bianchi's permutability theorem\cite{eisen} gives a new solution $\f_{h}$ by
\be
\f_{h} = \f_{g} + 4\mbox{tan}^{-1}({\d_{1} + \d_{2} \over \d_{1} - \d_{2} }
\tan{\f_{a}- \f_{b} \over 4 })
\ee

In this section, we show that all these properties generalize to the matrix
sine-Gordon theory.
Recall that the linear equation for the matrix sine-Gordon equation is
\be
[\pp + \gi \pp g -\l T ]\J_{g} = 0  \ ; \ [\pb + \li \gi \Tb g ] \J_{g} = 0
\ee
where $T, \Tb$ are as in Eq.(20). The BT between two solutions $g$ and $f$
of the matrix sine-Gordon equation may be defined in terms of $\J$,
\be
\J_{f} = {\l \over \l + i\d }(1 + {\d \over \l } f^{-1} \Tb g)\J_{g}
\ee
where $\d $ is a real B\"{a}cklund parameter. $\J_{f}$ and $\J_{g}$ both
satisfy
the linear equation with respect to $f$ and $g$. On the other hand, if Eq.(115)
is
combined with the linear equation (114) to elliminate $\J$, then we have an
equivalent
expression for the BT in terms of $f$ and $g$,
\be
f^{-1} \pp f - \gi \pp g + \d [ \gi \Tb f \ , \ T ] = 0
\ee
and
\be
gf^{-1} \Tb - \Tb g f^{-1} - \d \pb g\gi \Tb + \d \Tb \pb ff^{-1} =0
\ee
Also, the unitarity condition, $\J (\l ^*)\J^{\dagger }(\l ) = 1$, requires
that
\be
f^{-1} \Tb g - \gi \Tb f =0.
\ee
Eqs. (115) - (118) consitute the BT for the matrix sine-Gordon equation. With
$g\ , \ T \ , \ \Tb$
 as in Eqs.(20) and (21), Eqs.(116) and (117) in block components are
\ben
f^{-1}_{1}\pp f_{1} - \gi_{1}\pp g_{1} + \k \d (\gi_{2}f_{1} - \gi_{1}f_{2}
)&=&
0 \nonumber \\
f^{-1}_{2}\pp f_{2} - \gi_{2}\pp g_{2} - \k \d (\gi_{2}f_{1} - \gi_{1}f_{2}
)&=&0
\een
and
\ben
\pb f_{1}f^{-1}_{1} -\pb g_{2}\gi_{2} + {1\over \d }(g_{2}f^{-1}_{2} -
g_{1}f^{-1}_{1}) &=& 0 \nonumber \\
\pb f_{2}f^{-1}_{2} -\pb g_{1}\gi_{1} - {1\over \d }(g_{2}f^{-1}_{2} -
g_{1}f^{-1}_{1}) &=& 0,
\een
while Eq.(118) becomes
\be
f^{-1}_{1}g_{2} = \gi_{1}f_{2} .
\ee
Note that Eqs.(119) and (120) are consistent with the constraint equation (13).
We now show that the matrix sine-Gordon theory admit also a nonlinear
superposition
rule of solutions which generates a new solution by purely algebraic means.
This
is given by the permutability of the Bianchi diagram.
Let $a$ and $b$ are solutions of the matrix sine-Gordon equation generated by
 the BT from a known solution $g$  with the B\"{a}cklund
parameters $\d_1 $ and $\d_2$ respectively. Further, let $h$ and $h^{'}$ denote
solutions
obtained by applications of the BT with parameter $\d_2 $ to $a$ and with
parameter $\d_1$ to $b$. Then, the permutability of the Bianchi diagram
requires $h=h^{'}$.  In terms of the BT in Eq.(115), this means that
\ben
\J_{h} &=& {\l \over \l + i\d_2 } \ {\l \over \l + i\d_1 }(1+ {\d_{2}\over \l }
h^{-1}\Tb a)(1+{\d_{1}\over \l }a^{-1}\Tb g)\J_{g} \nonumber \\
&=& {\l \over \l + i\d_1 } \ {\l \over \l + i\d_2 }
(1+ {\d_{1}\over \l }h^{-1}\Tb b)(1+{\d_{2}\over \l }b^{-1}\Tb g)\J_{g}
\een
or
\be
(1+ {\d_{2}\over \l }h^{-1}\Tb a)(1+{\d_{1}\over \l }a^{-1}\Tb g)
=   (1+ {\d_{1}\over \l }h^{-1}\Tb b)(1+{\d_{2}\over \l }b^{-1}\Tb g )
\ee
which, when solved for $h$ using the relation $\Tb ga^{-1}\Tb^{-1} = ag^{-1}$,
gives
\be
h= g(\d_{1}b^{-1}-\d_{2}a^{-1})(\d_{1}a^{-1} - \d_{2}b^{-1})^{-1} .
\ee
or, in terms of $h_1$ and $h_2$,
\ben
h_{1 } &=&  g_{1} (\d_{1}b_{1}^{-1}-\d_{2}a_{1}^{-1})(\d_{1}a_{1}^{-1} -
\d_{2}b_{1}^{-1})^{-1}  \nonumber \\
h_{2 } &=&  g_{2} (\d_{1}b_{2}^{-1}-\d_{2}a_{2}^{-1})(\d_{1}a_{2}^{-1} -
\d_{2}b_{2}^{-1})^{-1}  \ .
\een
It is easy to check that $h$ is unitary if $a, b$ are unitary.
This is the nonlinear superposition rule of the matrix sine-Gordon equation
which allows one to generate a new solution from a known one by purely
algebraic means.
In the abelian limit, we may take $h=\exp{i\f_{h} \s_{3} /2 } \ ,  \
b= \exp{i\f_{b} \s_{3} /2}\ , \ a= \exp{i\f_{a} \s_{3} /2 }\ , \ g=
\exp{i\f_{g} \s_{3} /2 }$, then Eq.(124) reduces precisely to the nonlinear
superposition rule of the sine-Gordon equation  in Eq.(113).

Finally, we obtain one and two soliton solutions of the theory using the BT.
We take the trivial solution to be a vacuum given by $f_1 = f_2 = f$
for a constant $SU(2)$ matrix $f$. Then, the 1-soliton solution in terms of
$g_1$, $g_2$ is obtained through the BT in Eqs.(119) and (120) which,
after redefining  $g_{1} , g_{2}$ by
$  g_1 f^{-1} \rightarrow g_{1} ,\ \  g_2 f^{-1} \rightarrow g_2 $, becomes
\ben
g^{-1}_1 \pp g_1 - \k \d ( g^{-1}_2 - g^{-1}_1 ) &=& 0 \nonumber \\
\pb g_1 g^{-1}_1 + {1 \over \d} ( g_2 - g_1 ) &=& 0.
\een
and the same equation with $g_{1}$ and $g_2$ interchanged.
If we use the parametrization for $g_1$ and $g_2$,
\be
g_{1} = \pmatrix{ u  & i\sqrt{1-u\ub }e^{i\f } \cr
  i\sqrt{1-u\ub }e^{-i\f }& \ub } \ , \
g_{2} = \pmatrix{ v  & i\sqrt{1-v\vb  }e^{i\q } \cr
  i\sqrt{1-v\vb }e^{-i\q }& \vb  }  \ ,
\ee
 Eq.(126) resolves into
the component equations;
\ben
\ub  \pp u - u \pp \ub  - 2i(1-u \ub ) \pp \f- 2\k \d (\vb -\ub )
&=& 0 \nonumber \\[3mm]
\pp u - \k\d ( e^{i(\f -\q)} \sqrt{1-u \ub } \sqrt{1-v \vb } -1 +u \vb  )
&=& 0 \nonumber \\
u \pb \ub  - \ub  \pb u -2i(1-u \ub  )\pb \f+ 2 {1 \over \d} (u-v) &=& 0
\nonumber \\
\pb u - {1 \over \d} (e^{i(\q -\f )} \sqrt{1-u \ub }
\sqrt{1-v \vb } -1 + u\vb ) &=& 0
\een
and the same equation with the interchange; $u \leftrightarrow v ,  \f
\leftrightarrow \q $.   In addition,  the traceless conditon of Eq.(126)
requires that $u+\ub  = v+ \vb $.
The unitarity condition, Eq.(121), in this case requires that
$u=\vb $ and $\f= \q+\pi $. Then, equations for $u$ and $\ub $ become
\ben
({1 \over \k\d }\pp - \d \pb )u &=& 0 \nonumber \\
({1 \over \k\d }\pp + \d \pb )u &=& -4+2u \ub  + 2 u^2
\een
and the same equation with $u$ and $\ub $ interchanged.
This equation can be readily integrated
to give $u+\ub  = 2 \tanh (2 \D + \eta )\ ; \
\D = -\k\d z -  \zb /\d  $ and
\be
u = \tanh (2\D + \eta)  + i c {1 \over \cosh (2\D +\eta )},
\ee
where c is an arbitrary constant. In addition, when the solution $u$
is used, Eq.(128) can be solved for $\f $ such that $\f $ is a constant.
In terms of $g_1$ and $g_2$, this means that
\ben
g_{1} &=& g_{2}^{-1}= {1-N \over 1+N} \nonumber \\
N &=& ie^{2\D  +\eta } a_{k}\s_{k} \ , \
V \equiv ( 1 +  \k \d^2 )/ ( 1 - \k \d^2 )
\nonumber \\
\D &=& -\k\d z -{1\over \d } \zb  =
\pm {  \sqrt{-\k}  \over \sqrt{1 - V^2}}(x - Vt)
\een
where $\eta, a_{i}$ are arbitrary constants coming from $c$ and $\f $
with  normalization $a_{i}a_{i} =1$. This agrees precisely with the
1-soliton solution in Sec.3.

In order to obtain two soliton solutions, we may apply the nonlinear
superposition
rule to a couple of one soliton solutions obtained by the BT with parameters
$\d_1 $ and $\d_2 $ such that
\be
a_1 = a_2^{-1}= {1-N_1 \over 1+N_1}, \ ; \ b_1=b_{2}^{-1}= {1-N_2 \over 1+N_2}
\ee
where $N_1 , \ N_2$  are given in Eq.(131) with repective parameters
$\d_1 ,\eta_1  $ and $\d_2 , \eta_2 $.
Then, from Eq.(125) we obtain the 2-soliton solution,
\ben
h^{-1}_1 &=& \left( \d_1 {1+N_1 \over 1-N_1} -\d_2 {1+N_2 \over 1- N_2} \right)
\left( \d_1 {1+N_2 \over 1-N_2} -\d_2 {1+N_1 \over 1-N_1} \right)^{-1}
\nonumber \\
h^{-1}_2 &=& \left( \d_1 {1-N_1 \over 1+N_1} -\d_2 {1-N_2 \over 1+ N_2}
\right) \left( \d_1 {1-N_2 \over 1+N_2} -\d_2 {1-N_1 \over 1+N_1} \right)^{-1}
\ .
\een
It is now a straightforward but amusing exercise to check that
$h_1^{-1}$ is equal to $(A+ B^k_{+} \s_k )  / \det Y$ of Eqs.(79)
and (80) while
$h^{-1}_2$ is equal to $(A+ B^k_{-} \s_k )/ \det Y$.

\section{Discussion}
Throughout this Paper, we have analyzed various classical properties of the
matrix
sine-Gordon theory for the coupling constant $\k < 0$. For $\k > 0$,
the vacuum structure changes, i.e. degenerate vacuua occur at $\f = (2n+1)\pi $
for
integer $ n $ so that $ \gi_{1}g_{2}=-1$, or for $g_{1}=-g_{2}=f$ for
arbitrary $f$. This reflects the symmetry of the matrix sine-Gordon theory
under the exchange;
\be
g_{1} \leftrightarrow g_{1} \ , \
g_{2} \leftrightarrow - g_{2}   \ , \
\k \leftrightarrow -\k  \ .
\ee
Thus, the matrix sine-Gordon theory with $\k > 0$ is identical with the $\k <
0$ case
except the sign change of $g_2$. This type of symmetry arises due to a specific
choice of the potential term  and can be generalized to other types of
nonabelian
sine-Gordon theory considered in  \cite{park} \cite{shin} \cite{hol}.
 In the abelian sine-Gordon case, the potential is given by $\k \cos{\f}$ and
the
symmetry becomes the well-known one;
\be
\f \leftrightarrow \f + \pi \ , \
\k \leftrightarrow -\k  \ .
\ee
In finding exact solutions, solitons and breathers, we have assumed asymptotic
boundary
conditions $g_1 , g_2  \rightarrow 1$ as $ x \rightarrow \pm \infty $. However,
this
boundary condition may be relaxed to accomodate other types of solutions. For
example,
if we impose a periodic boundary condition, i.e. choose the underlying manifold
to be a torus, we can not
take a nonlocal gauge where $A=\bar{A}=0$  except for the trivial holonomy
sector
of the flat connection $A$ and $\bar{A}$. Nontrivial sectors could lead to
generalizations of solitons and breathers of the matrix sine-Gordon theory
characterized
by the nontrivial holonomy.
Another important classical aspect of the matrix
sine-Gordon theory which we have not considered is the hamiltonian struture
of the theory. As in the case of solitons, we expect a nontrivial
generalization  of the hamiltonian structure of the sine-Gordon  theory,
especially
a nonabelian generalzation of the R-matrix.

The quantum matrix sine-Gordon theory is equally important and provides a
nontrivial generalization of the quantum abelian sine-Gordon theory.
Moreover, it is a natural quantum field theory framework for the
Zamolodchikov's integrable $\Phi_{(2,1)}$ perturbation of Ising model.
In the parafermion case, the S-matrix obtained by Zamolodchikov
using the operator algebra has been explained nicely in terms of the
complex sine-Gordon theory\cite{hol2}. Since the matrix sine-Gordon theory
carries essentially the same structure as the complex sine-Gordon theory, one
expect
that a similar analysis is possible for this case using a semiclassical WKB
approximation. Work in this direction is in progress and will appear elsewhere.

\vglue .2in
{\bf ACKNOWLEDGEMENT}
\vglue .2in
 We would like to thank T.Hollowood for discussion.
 This work was supported in part by Research Fund of Kyunghee University,
 by the program of Basic Science Research, Ministry of Education BSRI-95,
 and by Korea Science and Engineering Foundation through CTP/SNU and the
Korea-Japan Cooperative Science Program.
\vglue .2in

\newpage
\begin{figure}
\vskip -0 cm
\centerline{\epsfxsize 4.5 truein \epsfbox {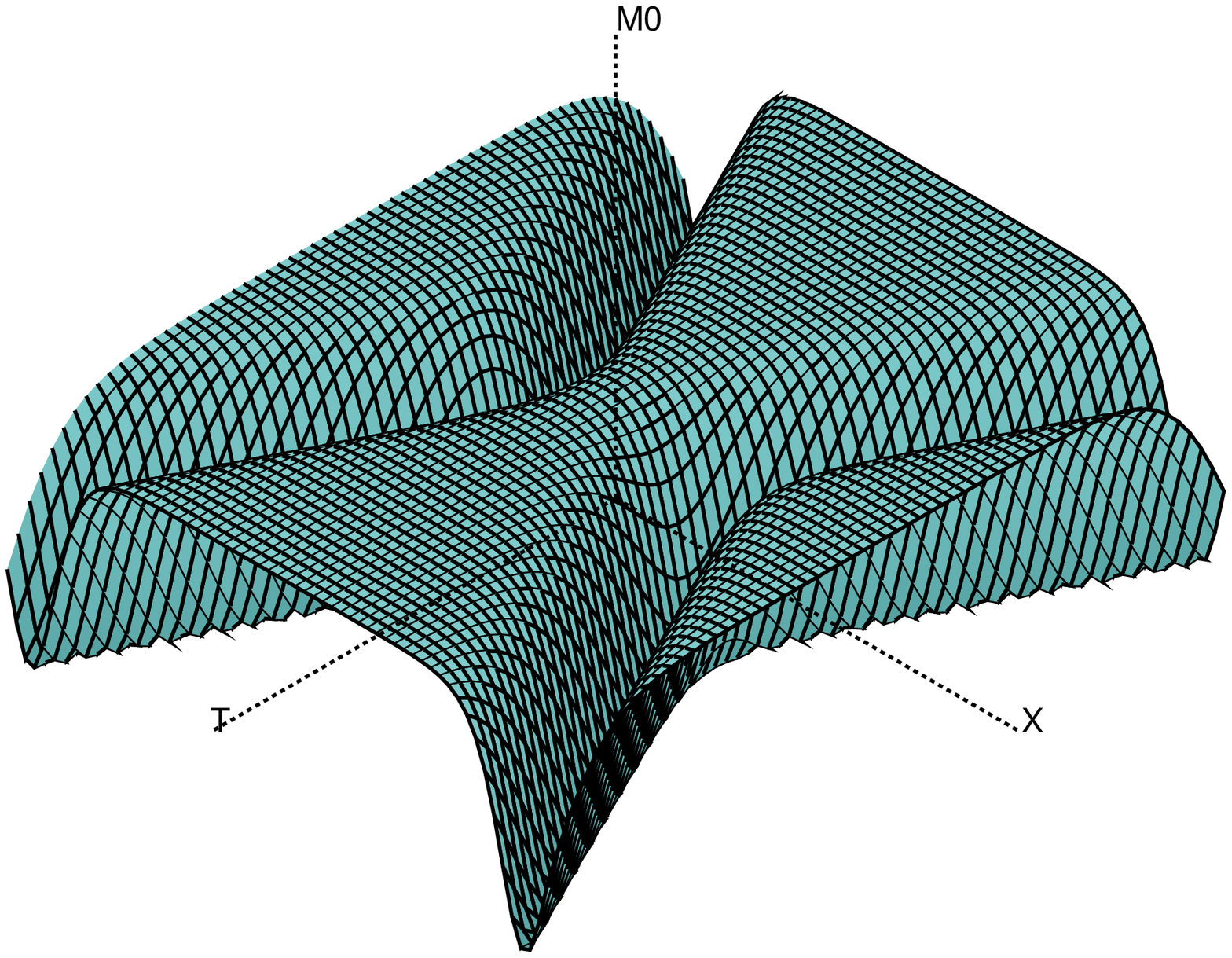}}
\nobreak
\vskip 0.5 cm\nobreak
\vskip .1 cm
\caption{ $M_{0} $ in soliton - soliton scattering}
\end{figure}

\begin{figure}
\vskip -0 cm
\centerline{\epsfxsize 4.5 truein \epsfbox {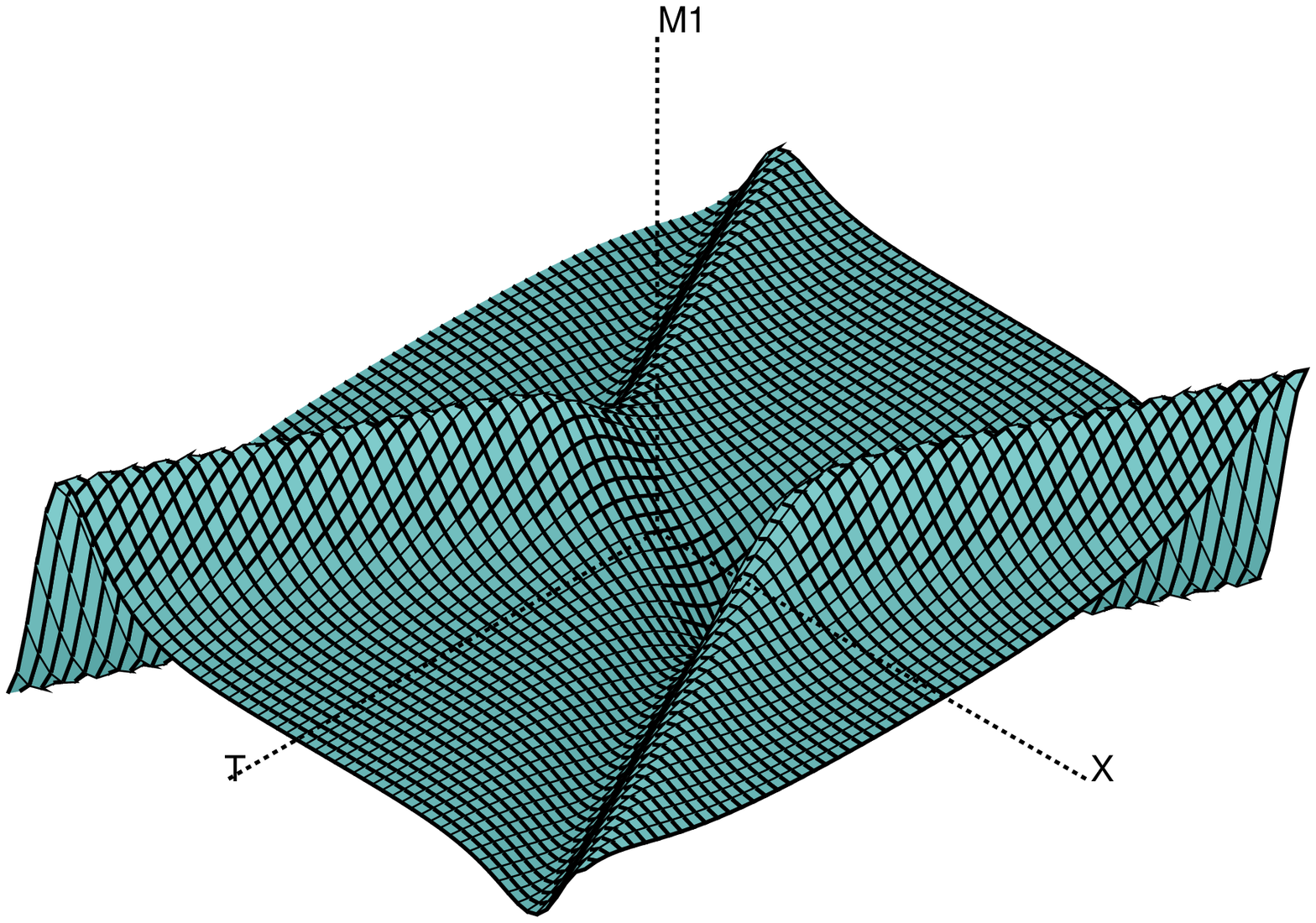}}
\nobreak
\vskip 0.5 cm\nobreak
\vskip .1 cm
\caption{ $M_{1}$ in soliton - soliton scattering }
\end{figure}

\newpage
\begin{figure}
\vskip -0 cm
\centerline{\epsfxsize 4.5 truein \epsfbox {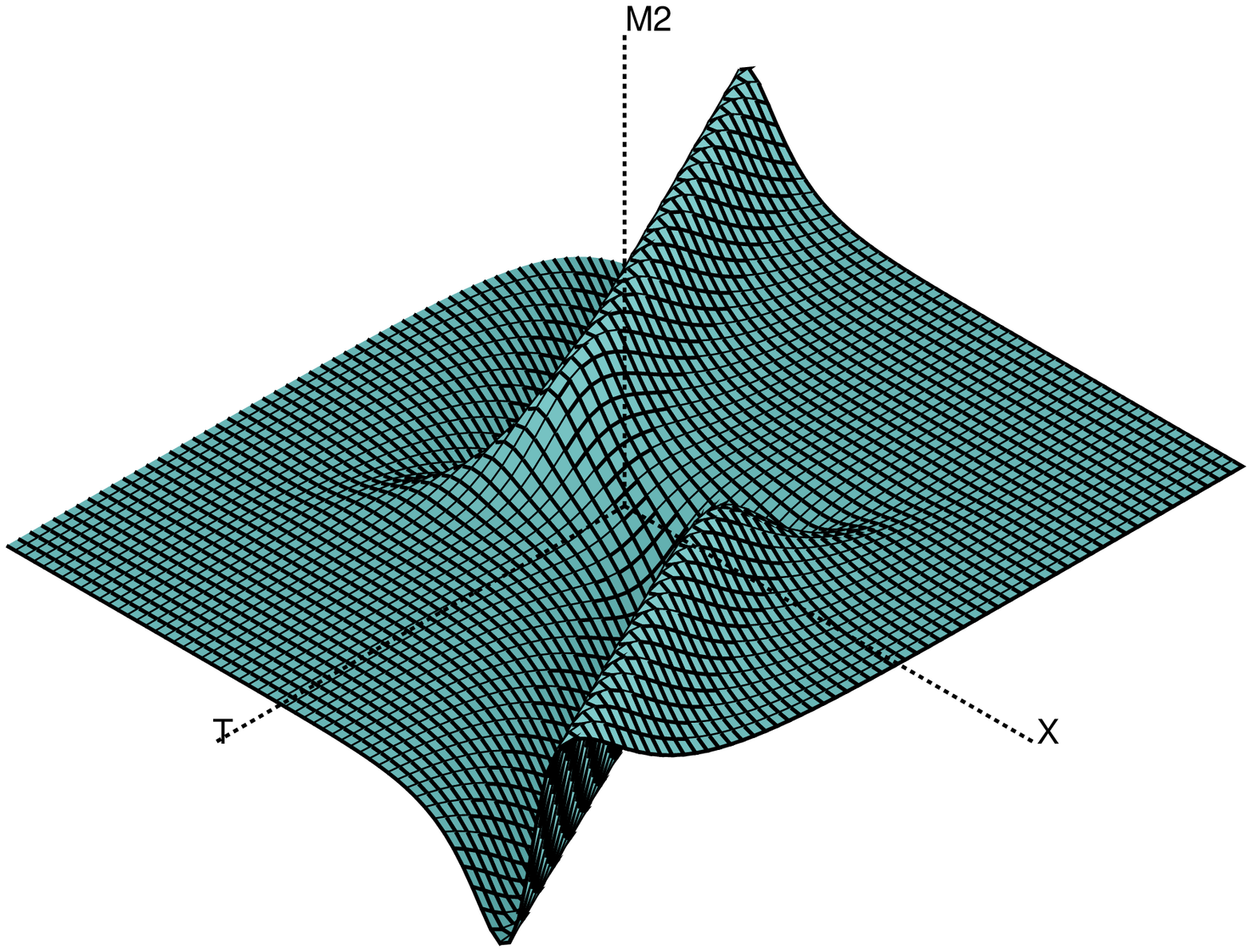}}
\nobreak
\vskip 0.5 cm\nobreak
\vskip .1 cm
\caption{$M_{2}$ in soliton - soliton scattering}
\end{figure}

\begin{figure}
\vskip -0 cm
\centerline{\epsfxsize 4.5 truein \epsfbox {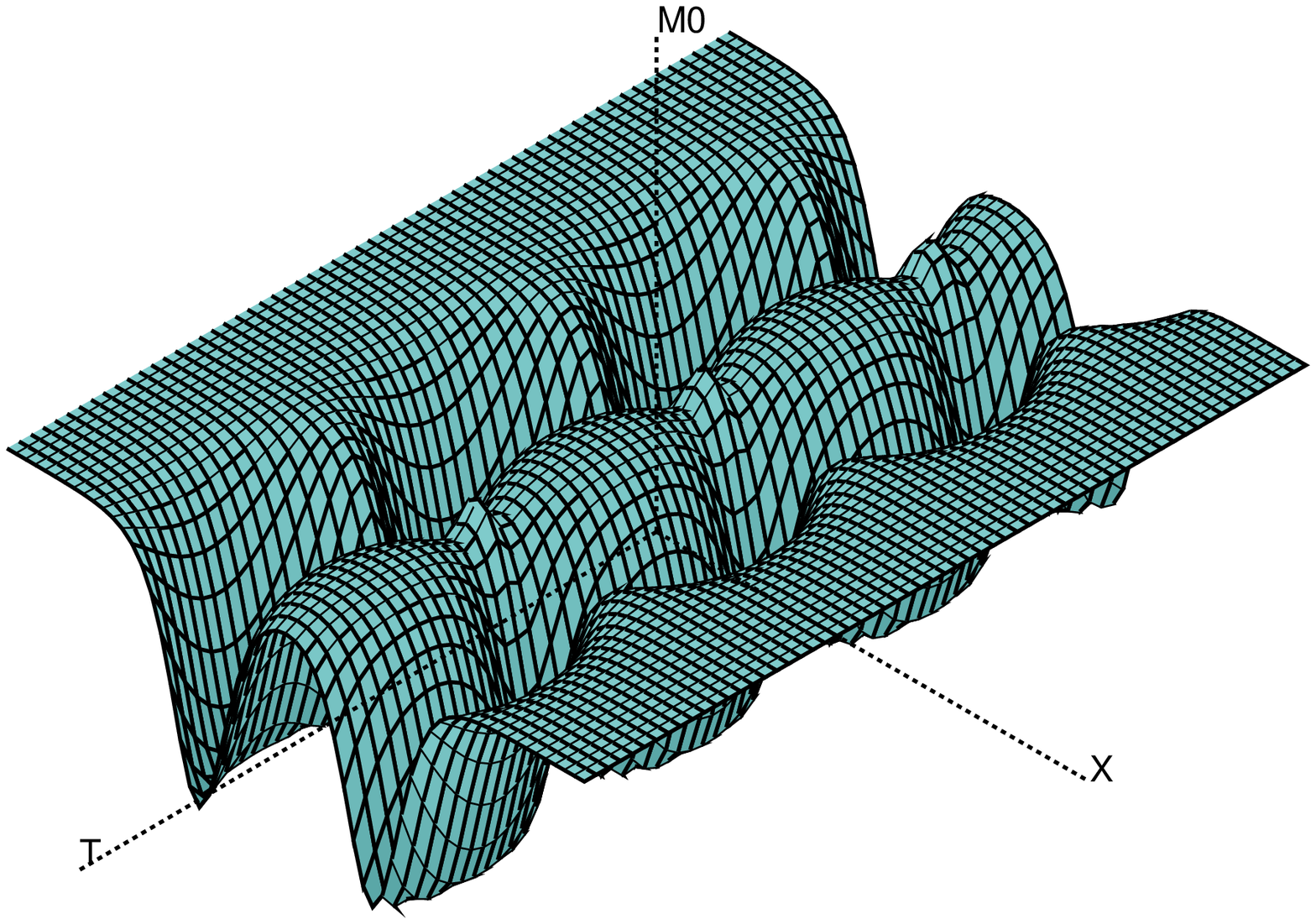}}
\nobreak
\vskip 0.5 cm\nobreak
\vskip .1 cm
\caption{$M_{0}$ of nonabelian breather }
\end{figure}

\newpage
\begin{figure}
\vskip -0 cm
\centerline{\epsfxsize 4.5 truein \epsfbox {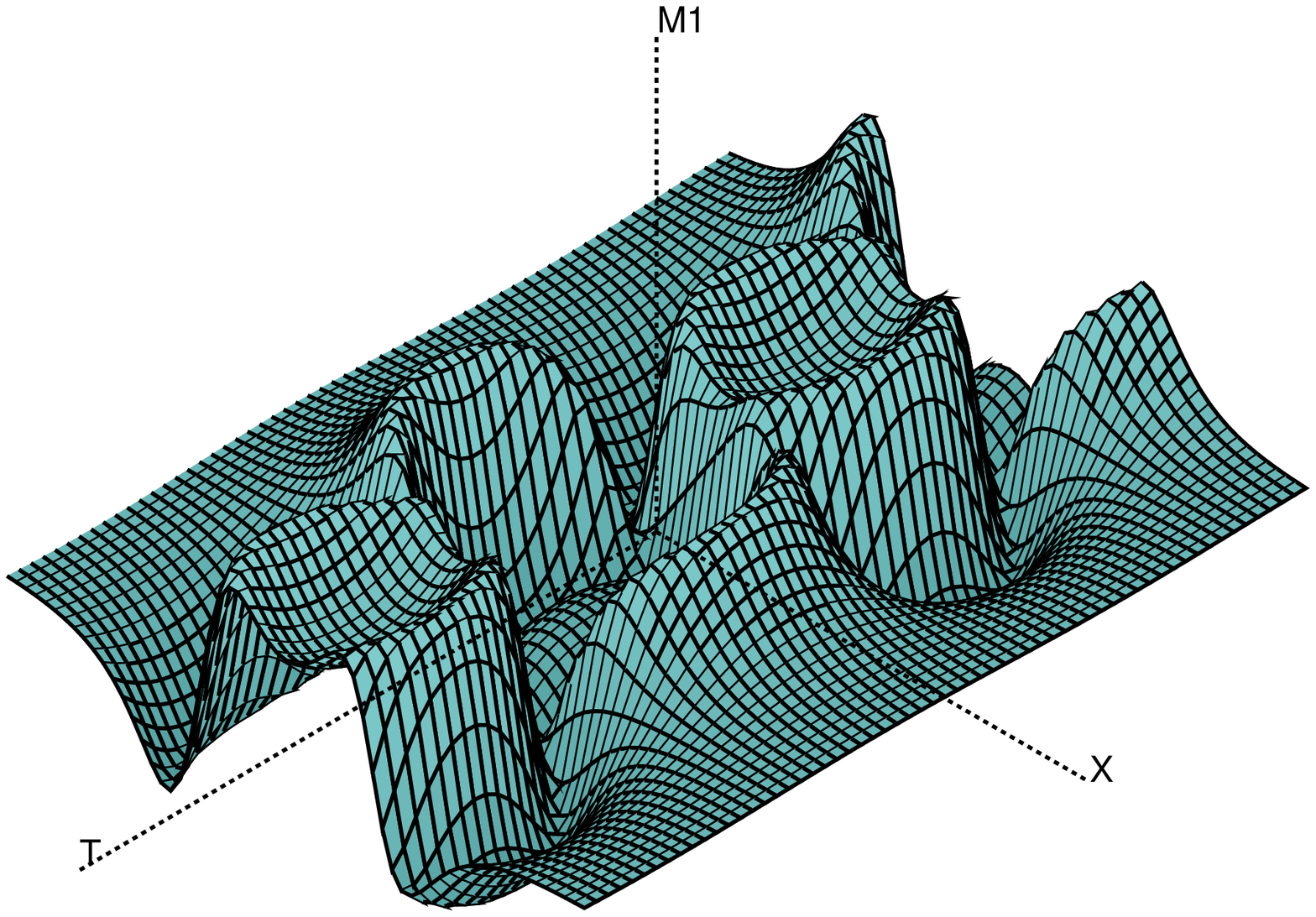}}
\nobreak
\vskip 0.5 cm\nobreak
\vskip .1 cm
\caption{ $M_{1}$ of nonabelian breather }
\end{figure}

\begin{figure}
\vskip -0 cm
\centerline{\epsfxsize 4.5 truein \epsfbox {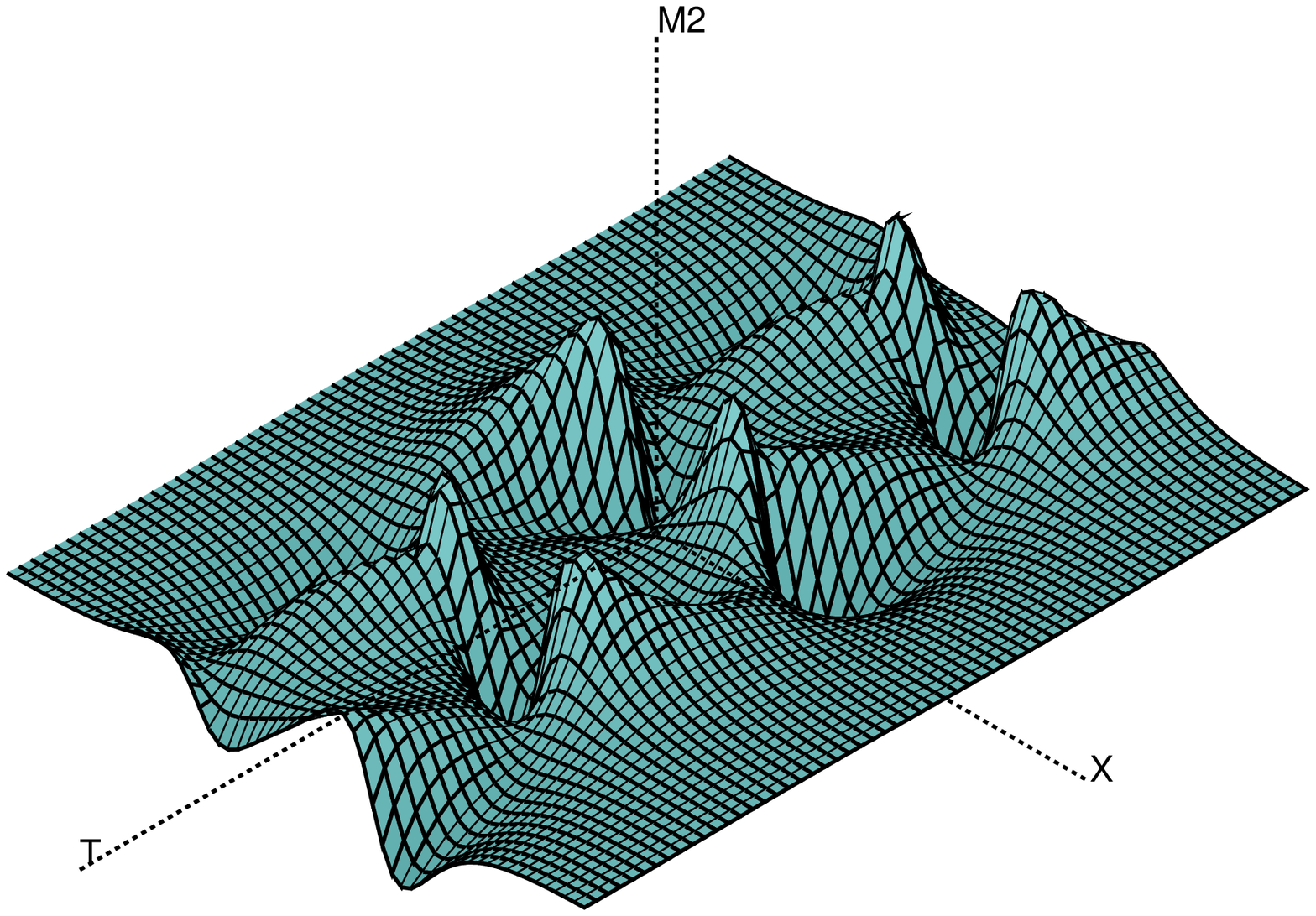}}
\nobreak
\vskip 0.5 cm\nobreak
\vskip .1 cm
\caption{$M_{2}$ of nonabelian breather}
\end{figure}

\end{document}